# Extended Distance-based Phylogenetic Analyses Applied to 3D *Homo* Fossil Skull Evolution

**Peter J. Waddell**[1]

pwaddell.new@gmail.com

[1]Ronin Institute, 1657 Upland Dr, Columbia, SC 29 204

.

This article shows how 3D geometric morphometric data can be analyzed using newly developed distance-based evolutionary tree inference methods, with extensions to planar graphs. Application of these methods to 3D representations of the skullcap (calvaria) of 13 diverse skulls in the genus *Homo*, ranging from *Homo erectus* (*ergaster*) at about 1.6 mya, all the way forward to modern humans, yields a remarkably clear phylogenetic tree. Various evolutionary hypotheses are tested. Results of these tests include rejection of the monophyly of *Homo heidelbergensis*, the Multi-Regional hypothesis, and the hypothesis that the unusual 12,000 year old (12kya) Iwo Eleru skull represents a modern human. Rather, by quantitative phylogenetic analyses the latter is seen to be an old (200-400kya) lineage that probably represents a novel African species, *Homo iwoelerueensis*. It diverged after the lineage leading to Neanderthals, and may have been driven to extinction in the last 10kya by modern humans, *Homo sapiens*, another African species of *Homo* that appeared about 100kya. Another enigmatic skull, Qafzeh 6 from the Middle East about 90kya, appears to be a hybrid of two thirds near, but not, anatomically modern human and one third of an archaic lineage diverging close to classic European Neanderthals. Overall, the tree clearly implies an accelerating rate of skullcap shape change, and by extension, change of the underlying brain, over the last 400kya in Africa. This acceleration may have extended right up to the origin of modern humans. Methods of distance-based evolutionary tree inference are refined and extended, with particular attention to diagnosing the model and achieving a better fit. This includes power transformations of the input data which favor root Procrustes distances.

# 1 Introduction

This is a worked example of the types of phylogenetic distance analysis that are possible using pairwise distances and least squares fitting. It falls within the category of maximum likelihood analyses of pairwise distance data where the model being considered is a tree. It is an example of a general linear model, particularly when the distances are subject to a transformation in the hope of making them better fit the model at hand. It covers and extends the types of methodology that have been used for distance-based phylogenetic analyses previously (e.g., see reviews by Swofford et al. 1996, Felsenstein 2004). This includes recent developments such as NeighborNet (Bryant and Moulton 2004), residual resampling (Waddell et al. 2011) and flexibly weighted least squares (Waddell et al. 2010). These methods as implemented in PAUP*4.0 (Swofford 2000) and SplitsTree4 (Huson and Bryant 2006), are combined with detailed statistical diagnosis based on PERL programs and worked examples in Excel.

The type of data chosen to illustrate this work is 3D morphometric data. Part of the reason for this is that while geometric morphometrics has rapidly advanced over the past two decades, the analysis of such data in a phylogenetic context has seriously lagged (Adams et al. 2004, 2013). This is despite interesting contributions such as Couette et al. (2005). The data chosen to work with here are pairwise distances based on the differences in 3D measurements of fossil human and pre-human skulls (Havarti et al. 2011, 2013). That is, the changes in shape, but not directly of size, of the bones at the top of the skull that encase the top of the brain (the skullcap or calvaria) in members of the genus *Homo* (Klein 2009). The shape of these bones, the frontals, parietals and occipital squama (the part of the occipital behind the top of the spine) are represented by many of their margins and sutures. Skulls generally represent the great majority, probably over 90%, of the useful morphological phylogenetic evidence we have for the evolution of our genus *Homo* over the past 2 million years (Schwartz et al. 2004). The shape of the skullcap, in turn, constitutes most of the best evidence for general changes in brain size and shape. In some cases, the inside of the skull can even yield an endocast with surface features of the brain visible, and this detailed information poses future challenges for shape quantification and digitalization (Schwartz et al. 2005, Klein 2009). The ~100 3D measurements of the shapes of these bones are converted into a distances matrix via the pairwise Procrustes procedure (Zelditch et al. 2004, Adams et al. 2013). The Procrustes procedure is a key method of the growing and relatively new field of geometric morphometrics, whereby the 3D shape of objects is quantified and analyzed.

While skulls represent most of the phylogenetic information on the evolution of *Homo*, and while geometric morphometrics is at the forefront of many fields of biology, anatomy, and paleontology, the methodology on how to analyze such data within a phylogenetic or evolutionary tree context, is very sparse and little expanded over the past couple of decades (Adams et al. 2013). This is becoming a particularly critical deficiency, not least since genomes are beginning to stretch deeper into time and reveal specific details of lineages within the genus *Homo* (Krause et al. 2011, Reich et al. 2011) and their genetic interactions such as introgression or interbreeding (Green et al. 2010, Waddell et al. 2011b, Waddell and Tan 2012, Waddell 2013). In turn, what genomic studies hypothesize these DNA samples to represent, is often then embedded back into a morphological framework of the evolution of the genus *Homo*. This morphological framework in turn remains very contentious and ambiguous in terms of specific affiliations, as there is no generally accepted methodology for inferring evolutionary lineages (e.g. Schwartz et al. 2013). There have been lots of quantitative multivariate methods used, such as Principal Components Analysis (e.g., Zelditch et al. 2004, Adams et al. 2013), but these are not phylogenetic methods. They do not directly address the critical evolutionary question of lines of descent and interrelationships, probably the key point of topical concern in the evolution of archaic *Homo* (e.g., Spoor 2013, Schwartz et al. 2013). Thus, the quantitative phylogenetic analysis of morphological change in the genus *Homo*, which is largely in the shape of the skull, is critical to progress in understanding human/pre-human evolution and speciation.

One consequence of this lack of adequate methodology is that even the question of what "species" existed is highly contentious. In studies of human evolution there is a strong desire to shoehorn most descriptions of the last two million years of human evolution into 2-5 species (Lordkipanidze et al. 2013, Schwartz et al. 2013, Bermúdez de Castro et al. 2014) or even as few as one (Wolpoff 1999). This is even more problematic because species concepts are vague and what biologists make of them obviously do not define what the real lines of descent were, whereas the direct study of evolutionary relationships is directly critical to our understanding what species are and were in the context of the genus *Homo* (Schwartz et al. 2005).

Some terminology is due here. In this article the term "human" is restricted to members of the species *Homo sapiens*, which is the clade of living members of the genus *Hom*o and all descendants of their last genetically dominant common ancestral population. This population probably lived in Southern Africa about 100kya (Penny et al. 1995, Waddell and Penny 1996). Thus terms like *Homo sapiens sapiens*, or modern human, mean the same thing. Anatomically modern human means cannot be distinguished from a range of typical modern humans based on anatomy, which for paleontology, usually means bones. Other members of the genus *Homo*, that would seem to be different species, not least since they were apparently competitively replaced with minimal genetic exchange, are described generally as hominins, pre-humans, other members of the genus *Homo*, or archaic hominins. (There is, unfortunately, not a readily available general term for "members of the genus *Homo"*, as there is for australopithecines; the term "homos" has other connotations).

Key to phylogenetic analyses based on distances is useful software. One purpose of this article is to provide worked examples for potential implementation in the widely used phylogenetic program, PAUP* (Swofford 2000). This program has an extensive range of tree search strategies and criteria, as well as tree manipulation and diagnostic methods. These include a wide range of flexibly weighted least squares methods (fWLS), which act as maximum likelihood methods when the input data are distances. The extension of statistical tests to compared observed and expected data, to distances in addition to discrete characters, is important, not least since some data sets, such as this, are presently only really amenable to distance-based analyses.

<u>The data used</u>. Specifically a set of 13 taxa was chosen to span the 2 million year period of *Homo* evolution and in particular to explore the affinities of a enigmatic skull from Iwo Eleru in Nigeria, West Africa (Brothwell and Shaw 1971, Shaw and Daniels 1984, recently reviewed by Allsworth-Jones et al. 2010). Despite being dated to only ~12-16 kya (Brothwell and Shaw 1984, Havarti et al. 2012), it seems to fall outside the variability of modern human skulls (which have a common ancestor ~100kya, Waddell and Penny 1996) and is closest to the skull LH18 from East Africa (Havarti et al. 2012) that is dated to ~130 to 490 kya (Millard 2008). To date it has been analyzed using non-phylogenetic methods that suggest it falls outside the range of variation of modern humans. However, if the distance measured between skull's shape is an additive accumulation of many genetic changes, then like sequence distances between full genomes, that set of taxa is expected to fit best onto a tree matching the lines of descent of each individual in the analysis. This has yet to be assessed, and should yield improved insights to this exciting find.

Other well-known skulls included in the analysis, working from the past towards the present, are two famous *Homo erectus/ergaster* skulls from East Africa (3733 and 3833 from roughly 1.6 million years ago = 1.6mya), that serve to root the tree (Klein 2009). Then there is the stunning, morphologically primitive, exceptionally well-preserved but very poorly dated Broken Hill/Rhodesia (now Kabwe/Zambia) skull (anywhere from ~0.4 to 1.8 mya, Millard 2008). This is the type specimen for *Homo rhodesiensis*. At about 200kya is the enigmatic Dali skull from China. Many would like to shoehorn both these two skulls into a poorly defined "species" called *Homo heidelbergensis* (e.g., Lieberman and Bar-Yosef 2005, Mounier et al. 2011). The latters' type specimen is a jaw (from Heidelberg), so it might be more appropriate to say they have been "jaw-boned" into this association. Whatever heidelbergensis is, it seems to artificially fill out our lack of understanding of specific lineages over a wide period of time from nearly 1 mya, right down to as recently as 200kya. It is a slightly more precise definition of what used to be called "archaic *Homo sapiens*", principally, although not always, with the exclusion of African fossils less than ~400kya. Looking at it critically, perhaps its great scientific contribution is to make us aware that evolution before distinct later forms such as Neanderthals, but well after classic erectus/ergaster, had not stopped. There was something of a trajectory of increased brain size in this period (Lieberman and Bar-Yosef 2005). Neanderthals separated off from modern humans perhaps 0.4 to 0.5 mya ago according to genetic evidence (Krause et al. 2011), and are represented here by two classic late European Neanderthals, "The "Old Man" of La Chapelle-aux-Saints and La Quina 5, both French and roughly 50 to 60,000 years old = 50-60 kya, or only ~0.05 mya). From West Africa we have Iwo Eleru and from East Africa, LH18 (Laetoli Hominin 18). While LH18 was originally recognized as somehow more "*Homo sapiens*" than Neanderthal (Magori and Day 1983), Kabwe, for example, was also considered "*Homo sapiens*", thus the concept of "*Homo sapiens*" at that time was fairly close to "Not classic *Homo erectus* nor classic Neanderthal". Its age was originally ascertained to be close to 120 kya, but now is suspected to lie in the range 120 to 490kya. Another interesting skull is Qafzeh6, probably a male from Palestine/Israel and about 90 kya. There were also Neanderthal forms in this area before and after this skull, which is clearly near modern in form, but lacks a modern chin, perhaps the best synapomorphy yet for humans. There are also two skulls from Europe at about 30kya, two of the famous Cro-Magnon skulls that probably represent descendants of a first, but increasingly seen as part of a second, wave of modern hunter gathers into Europe (and if a second wave, possibly with substantial genetic turnover). CroMagnon 1 is universally regarded as a male, and gives rise to much of the stereotype of a heavily built "Cro Magnon" man. Finally, there are two very recent KhoiSan skulls (one male and one female), which while recent themselves, this lineage intersects with that of the Cro-Magnons ~80 to 120 kya and can be considered to encompass most modern human skull variation in overall shape (in phylogenetic or evolutionary tree sense). For more details on these archaic skulls see Schwartz et al. (2005).

## 2 Materials and Methods

The data used here are a set of 3D morphometric measurements based on 19 landmarks and 78 semi-landmarks from a set of modern and fossil skulls of the genus *Homo* (Havarti et al. 2011). The data largely follow 3D points on the curves of the major sutures and spatial bone boundaries on the top of the skull, these being the mid-line (sagittal suture and continuation into the occipital to the inion), across the top of the brow (supraorbital ridge), the suture that runs from the near the top of the skull down the sides towards the top of the temple region and marks the boundary of the frontal and the parietal bones (coronal suture) and finally, the boundary of the parietal bones and the occipital squama at the rear of the skull (lambdoid suture). From the 3D coordinates, pairwise Procrustes distances were calculated using a variety of software, which agreed closely with the numbers from Havarti et al. (2011). Procrustes distances are the minimum possible unweighted least squares distance between the corresponding landmarks and semi landmarks of one skull to the other achieved by rescaling, translating and rotating. In this particular case, the square root of the sum of squares distance was primarily used as it fitted trees much better than the sum of squares. The distance data matrix used is given in relaxed PHYLIP format in the appendix.

A number of programs and scripts were used for these distance analyses. These include PAUP*, which is now at alpha version 136, a major prerelease of the next version (Swofford 2000). It incorporates a range of flexibly weighted least squares models (Waddell et al. 2010). Also used was the program SplitsTree (Huson and Bryant 2006) which enables the NeighborNet algorithm (Bryant and Moulton 2004). In addition, a variety of Perl scripts were used, which enabled routines such as residual resampling and output of a trees design matrix, from Waddell et al. (2011a) and Waddell and Tan (2012). More specific calculations were made in Excel, including its numerical optimizer, "Solver".

# 3 Results

## 3.1 Ordinary Least Squares And Treeness

A good first criterion for a tree search with distances is ordinary or unweighted least squares, with edge lengths constrained to be non-negative (OLS+, Felsenstein 1989, Swofford et al. 1996). Applying this criterion in PAUP* then doing a branch and bound search guaranteed to find the optimal tree (Swofford 2000) yields the weighted tree shown in figure 1. Multiple lineages, often grouped together, appear resolved and the rate of evolution is quite variable. The tree is drawn using an option to ladderize internal nodes down the page, which reveals a striking impression of "a great chain of being" within *Homo*. Should the reader's skull measurements be included, they are expected to fall closest to the three skulls in the lower right. This is striking result, as it clearly implies that the rate of evolution of the shape of this part of the skull, and by extension, the brain, has increased through time in the lineage leading to modern humans.

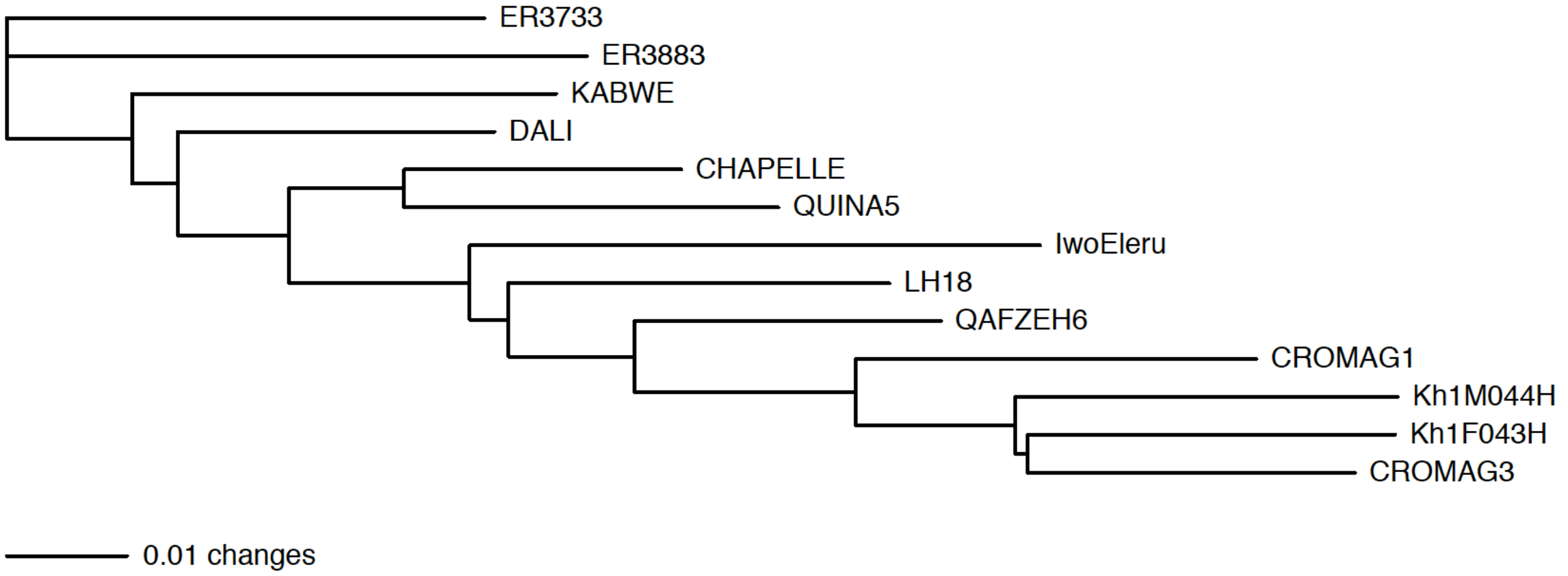

Figure 1. The OLS+ tree obtained by a branch and bound search. The ordering of edges in this tree, reading down the page, is used to refer to the taxa. The total residual sum of squares is 0.00217 while g%SD = 6.161 [$k$ = 24]. The skulls, with their probable sex, M = male, F = female and U = unclear, and their approximate ages in tens of kya, are ER3788 (Unclear 160 kya), ER3883 (U 160), Kabwe (M 49-178), Dali (U 180-230), Chapelle (M 5), Quina5 (F 6), Iwo Eleru (U 1), LH18 (U 13-49), Qafzeh6 (M 9), CroMag1 (M 3), CroMag3 (U 3), Kh1M044 (M 0), and Kh1F043 (F 0) (dates from Schwartz et al. 2004 and Millard 2008 in particular). The last four specimens are all typically assigned to *Homo sapiens sapiens*, or fully modern humans, yet CroMag1 falls outside the range of variation of most recent skulls. More importantly, in phylogenetic analyses (not shown) it nearly always diverges before a wide range of modern skulls (approximately 200, from all over the world, results not shown).

Without needing to place too much reliance on reconstructions of the ages of internal nodes, it is clear that evolution has sped up on the lineage to modern humans. The root node is at ~1.6 mya (Schwartz et al. 2004), that of the node at the modern human/Neanderthal split is ~0.4-0.5 mya (Krause et al. 2011), and the internal node leading to the modern humans is ~0.1mya (Waddell and Penny 1996). Further, Chapelle, Quina 5, Iwo Eleru and all the modern humans, are all in the range of only 0 to 0.06mya, clearly show this increasing evolutionary rate. It is also clear that most of the evolution in shape of the upper cranium has occurred since the divergence with Neanderthals, and does not show a clear sign of arrest, even with forms such as Qafzeh 6, which some researchers consider as being of essentially modern morphology (reviewed in Schwartz et al. 2004). Here it is seen as markedly less derived than modern human shapes and, in morphological terms, only about half to two thirds of the way derived from our distant ancestors with Neanderthals towards the modern condition. This is all the more true when it is recognized that CroMag1 is a particularly deeply diverging, and highly atypical in this regard, for a modern human skull. Thus, a lot of the shape change that suggests brain reorganization and perhaps new functions, rather than just increase in size, is a characteristic of the modern human lineage in particular.

Given that phylogenetic analysis of Procrustes distances is rarely done, and not yet in a statistical framework, is very important to assess in multiple ways whether the results are probably reliable. One way to approach this is to use PAUP* to evaluate the input distance data on random equally likely binary trees and use the results to gauge the "informativeness" or "treeness" of the data. PAUP* produced the histogram in figure 2b, along with four statistics: the mean, standard deviation, *g*1 and *g*2. The latter two statistics measure the third and fourth moments, that is, skew and kurtosis. In the case of discrete characters and parsimony, a g1 skew statistic of < -0.1 has been argued to indicate considerable "resolved tree" structure in the data (Hillis and Huelsenbeck 1992). From these new results, it seems probable the arguments made by Hillis and Huelsenbeck are equally applicable to the fit of distances to a tree. Here, with this data, the *g*1 skewness statistic of -2.62 seems to indicate a very large amount of resolved tree structure or "treeness" in the data. This is consistent with other evaluations made latter, including the ig%SD statistic, various $R^2$ statistics, and the residual resampling results. Figure 2a can be considered "zooming in" precisely on the upper part of the distribution portrayed in figure 2b. It is obtained by using branch and bound to guarantee finding all the near optimal trees. This is encouraging as many readers must doubt that these distances in shape would contain much useful phylogenetic information.

With this data there are just three trees that are very close to the optimal fit, which in this case are alternatives about the shortest internal edge seen in figure 1a. Following peaks in this distribution relate to the short internal edge adjacent to where Iwo Eleru joins the tree, and also where Dali joins the tree.

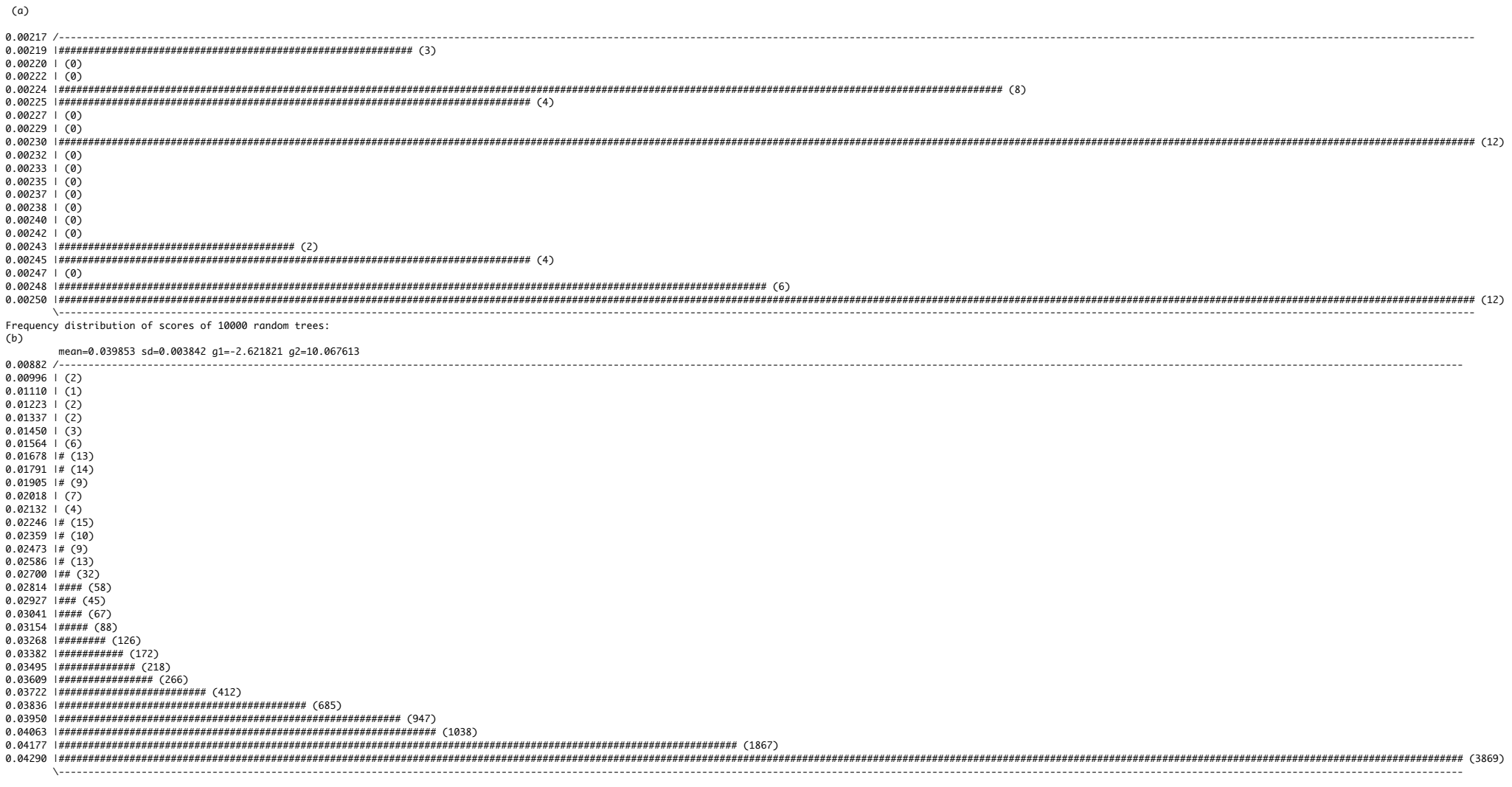

Figure 2. (a) The fit distribution of all trees better fitting than a sum of squares of 0.0025 as found using branch and bound. Graphic copied directly from the PAUP* log file. (b) Fit of 10,000 random trees, with g1 = -2.62 and g2 = -10.07.

### 3.2 Other distance based criteria

PAUP* now offers a range of least squares methods of distance analysis, some of the most flexible of which are the flexi-Weighted Least Squares (fWLS) methods (Waddell et al. 2010). These currently offer weights in two main forms. These are power weights of the form $w_i = \left(D_{obs_i}\right)^P$ with $P$ the free parameter (Felsenstein 1989), and exponential weights of the form $w_i = \exp\left(P' \times D_{obs_i}\right)$ , with $P$' as the free parameter (Waddell et al. 2010). Note the special case of $P$ = $P$' = 0 is the same as ordinary least squares, since all weights are then 1. In both cases, the optimal value of $P$ or $P$' for the data at hand can be estimated by finding the value that maximizes the log likelihood of the data under these models (Waddell et al. 2007, Waddell et al. 2010).

The results of fitting power and exponential weights to this data are shown in figure 3, as output by PAUP*4a136 (Swofford 2000). In this example, both exponential and power weights give very similar fit near their optima that both appear to be unique. The optimal fit is also very close to that of $P$ = $P$' = 0. The fit statistics reported here are, SS, the sum of squares, %SD, a percentage of misfit per distance onto the tree, g%SD, a percentage misfit that uses a geometric average of the weights and is inversely monotonic with the likelihood of the data for these models, $k$, the number of free parameters estimated, arith_mean, the arithmetic mean of the informative or non-diagonal distances, geo_mean, the geometric mean of the distances, IBLP, the internal branch length proportion or the sum of the internal edge lengths over the sum of all edge lengths, i%SD the internal %SD fit statistics, or %SD/IBLP, and, finally, ig%SD or g%SD/IBLP.

```
Distance settings:
  Objective function = unweighted least squares (power=0)
  Branch lengths constrained to nonnegative values
  Distance measure = user-input

Time used to compute distance scores = 0.01 sec (CPU time = 0.00 sec)

Distance scores for trees in memory:

Tree #            1
-------------------
SS          0.00217
%SD         5.90151
g%SD        6.10517
k                23
arith_mean  0.10645
geom_mean   0.10290
IBLP        0.17966
i%SD        32.8479
ig%SD       33.9815

Plot of %gSD against weight P ('*') and corresponding exponential weight ('+') for tree 1:

    Optimal P  = -1.01596 (%gSD = 6.0708)
    Optimal P' = -12.6409 (%gSD = 6.04153)

  8.3 +
      |
  8.1 +
      |+
  7.9 + +
      |
  7.7 +* +
      |
  7.5 + * +
      |  * +
  7.3 +   * +
      |    *                                                     ++
  7.1 +     *+                                                 ++
      |      *+                                               +
  6.9 +       *+                                            ++   **
      |        **                                         ++   **
  6.7 +          *                                      ++  ***
      |           **                                  ++  **
  6.5 +             *                               ++ ***
      |              **                           ++***
  6.3 +                ***                     ++***
      |                  +***               *****
  6.1 +                     +***************
      |                          ++++
  5.9 +----+----+----+----+----+----+----+----+----+----+----+----+
    -7.0 -6.0 -5.0 -4.0 -3.0 -2.0 -1.0  0.0  1.0  2.0  3.0  4.0  5.0

       (range for P' = -70.2738 to 50.1955)
```

Figure 3. Diagnostic output for the tree and data of figure 1. For these calculations, PAUP* first estimated the optimal value for $P$ (the power weights model) and $P$' (the exponential weights model) with respect to g%SD on the OLS+ tree. It then rounds the optimal $P$ value to the nearest integer and plots the g%SD value for the range of $P$ plus or minus six of this value, in increments of 0.2. For the exponential weights model, for each value of $P$ plotted, the value of $P$' which will give exactly the same ratio of the largest to the smallest weight is calculated, and the g%SD value at this $P$' value is plotted (Waddell et al. 2010).

The %SD and ig%SD statistics are intended to function in an analogous way to measures such as the consistency index (CI) or the retention index (RI), as used on parsimony trees (Swofford et al. 1996). That is, as general measures of fit that are hopefully roughly comparable across data sets, like the coefficient of determination, $R^2$, is in linear regression. The g%SD statistic is inversely monotonic with the likelihood (and, therefore, also the log likelihood) under these models, with the same nominal value of $k$ used, and on a fixed set of data. In calculating both of them, the weighted sum of squares is divided by an estimate of the degrees of freedom in the data, which is the number of distances $N$ minus $k$. If the only parameters fitted are the edge lengths and the sample variance for calculating the likelihood, then d.f. = $N$-$e_n$-1, where $e_n$ is the number of non-zero edge lengths. This is done to adjust for the shrinkage of the expected sum of squares due to freely fitting parameters, and it is expected this will overcome some of the tendency of g%SD to increase with adding taxa, analogous to the problem that CI and RI indices trend to markedly decrease as taxa are added to the tree. Of these statistics, it is the ig%SD statistic that may attract most attention, since it measures the noise to signal ratio in the model with respect to the distances between the most external internal nodes. Thus it helps to gauge how well the internal edges are being resolved. At around 34%, in this example, it suggests that the signal to noise ratio of each distance is still reasonable with regard to the size of the internal edges, consistent with the results of the g1 statistic in the randomization test performed above.

Also available in PAUP are minimum evolution (ME) based on least squares parameter estimation (ME/OLS) and balanced ME (BME) criteria, as well as three fast heuristic methods, UPGMA, NJ and BioNJ (Swofford et al. 1996, Gascuel and Steel 2006). The last two heuristics are close to a star-decomposition search using the BME criterion. BME uses a weighting system that is exponential, but it is only sensitive to how many internal nodes a distance crosses, not the size of the distance. For a binary tree, BME weights are of the form $w_i = 2^{in}$, where $in$ is the number of internal nodes crossed (Gascuel and Steel 2006). The optimal ME/OLS tree for this data is the same as the OLS+ tree, which is also the same as the tree obtained with the optimal $P$ and $P$' values for this data. The BME, NJ and BioNJ trees are all the same, but slightly different, from the OLS+ tree in that they all group LH18 and Iwo Eleru together as sister taxa. The g%SD (with $k$ = 24) for the optimal BME tree is 6.821. Later, it is shown how to make an assessment of how this compares to the flexi-weighted least squares trees.

Finally, the UPGMA methods assumes that the distances are no only additive, but all are additive on a clock-like tree (that is, ultrametric, Swofford et al. 1996). This is clearly not the case here, and the tree is radically different to that obtained with OLS+. Despite the data often not fitting well with UPGMA's expectations, such a tree inference method is still quite widely used in some fields, sometimes for the more vague objective of clustering based on overall similarity, but sometimes interpreted as a useful phylogenetic tree for the data, when it is unlikely to be so. The rooted UPGMA tree is shown in figure 4. It serves an illustration of the importance of appropriately matching model assumptions to data when inferring phylogenetic trees.

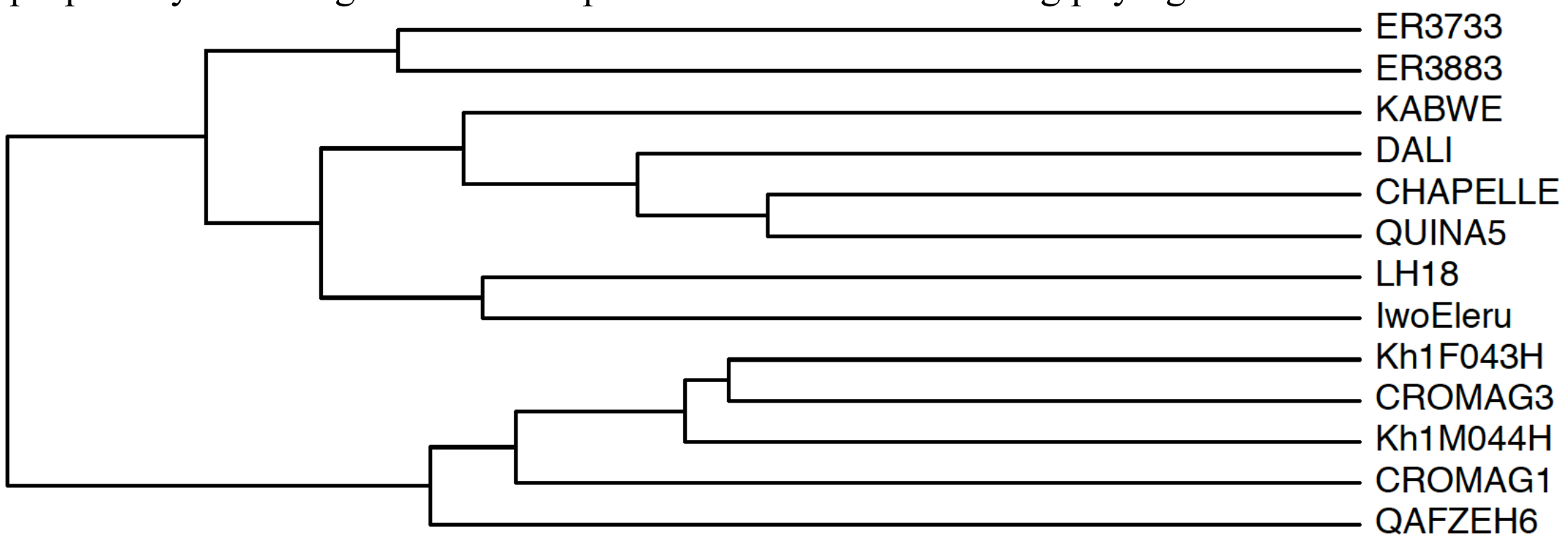

Figure 4. The UPGMA tree of the root Procrustes distances. Note the number of different clades generated compared to figure 1. When rates of evolution are uneven and the data are not sufficiently close to ultrametric, then UPGMA can produce very misleading trees, as here, with a mixture of misleading local edge interchanges as well as misrooting and more divergent rearrangements.

### 3.3 Likelihood ratio tests of models

It is useful to ask if the differences in fit of data to tree are significantly different for OLS+, the flexi-weighted least squares models with *P* and *P*' optimized and also the BME-based methods. This can be achieved with likelihood ratio tests, since the ordinary least squares model is a special case of all these other models. That was already noted with regard to the fWLS models, but it is also the case with the BME models, as their variance estimates for distances also intersection with the OLS model. For BME and binary trees, the variance of a distance is assumed to be proportional to $2^{inP''}$, where *in* is the number of internal nodes that distance crosses, and *P*'' = 1. When *P*'' = 0, this model too reverts to OLS weights.

It is possible to perform a likelihood ratio test of the null hypothesis that *P* = *P*' = 0 is the true set of weights. The probability of rejecting $H_0$ when it is true will be set at p = 0,05. That is to compare the log likelihood of $H_0$: *P'* = 0 (g%SD = 6.161, *k* = 24) with $H_1$: *P*' is not equal to zero. The maximum likelihood is achieved when g%SD is minimized, here at the value *P*' = -12.64 (g%SD = 6.097, *k* = 24). Since the g%SD for a fixed *k* is inversely proportional to the likelihood per distance, the log likelihood difference of the two data sets is equal to *n*(*n*-1)/2(-1/2{ln[g%SD($H_1$)/100]-ln[g%SD($H_0$)/100]}) = *n*(*n*-1)/4 ln[g%SD($H_0$)/g%SD($H_1$)] = 78/4ln(6.161/6.097) = 0.407. Under the null hypothesis, asymptotically and without boundaries, the distribution of this log likelihood ratio statistic is expected to be $\chi_1^2/2$, that is, a scaled chi-square distribution with one degree of freedom (Ota et al. 2000). Since *P* and *P'* here are free to be positive or negative there is no apparent boundary, although under unusual circumstances if other parameters such as edge lengths go to their boundary (e.g., non-negativity constraints) as *P* or *P*' take on their optimal values, then an exact $\chi_1^2/2$ distribution may not be realized. Since the actual $\ln L$ difference, which is much less than the critical value for $p$ = 0.05 ($\chi_1^2/2$ ~ 3.84/2), we fail to reject $H_0$ in favor of $H_1$.

Repeating the above test after a search across tree space at *P*' = -12.6409 sees a small change in the tree occur with the KhoiSan (Bushmen) male and female becoming sister taxa (also, potentially a bad case of gender confusion). The g%SD remains 6.097 under $H_1$ but increases to 6.164 for $H_0$, to give a log likelihood ratio of 0.426, which, again, is not significant. Note, it is wise to do this type of test both ways if a tree is not specified *a priori*, since there is a bias in favor of whichever hypothesis the tree search is made under.

Now the test is repeated for the BME model. The g%SD (*k* = 24) on the optimal BME tree is 6.821. For the OLS+ model under the same tree, the g%SD is 6.252 (*k* = 24). This test is being made is biased towards the BME model, since it is made on the optimal BME (not OLS+) tree and since under the OLS+ model, one of the internal edges (actually that linking Iwo Eleru and LH18) goes to the boundary with 0 length. Even so, the likelihood ratio statistic is strongly in favor of OLS+ at -3.397, thus we fail to reject $H_0$ again. The likelihood ratio difference is large enough, that with equal priors and in a Bayesian context, the relative posterior probability for each model would be nearly 30:1 in favor of OLS+. Thus the BME model and its favored tree are not considered strong competitors with the OLS+ model and its tree on this data.

### 3.4 The residual resampling robustness of the tree

In order to assess the robustness of different parts of the tree the residual resampling methods of Waddell and Azad (2009) seem particularly applicable, given a distance data set with no other information assumed. If it is assumed the data come from the model, then the only thing that needs to be specified is the variance. If the empirical variance estimate, $\hat{\sigma}^2$, is used, then for each expected distance in turn, a random sample is taken from *N*(0, $\hat{\sigma}^2$), and it is multiplied by the square root of the weight associated with that distance (for OLS always 1), and added to the expected distance. This is analogous to a sequence-based parametric resampling of sequences (Swofford et al. 1996), although an estimated variance is used. This tends to make the method much more conservative and robust, relative even to the better-known parametric bootstrap of trees in phylogenetics (Waddell and Azad 2009, Waddell et al. 2011a). These pseudo-replicate data sets are then analyzed in exactly the same way as the original data was analyzed, with the best tree(s) for each replicate saved. A typical way of showing support for a tree is, for each edge of the original tree count the fraction of the replicate trees that have this edge, or, alternatively calculate a consensus tree or graph of some sort, and do the same (Swofford et al. 1996). A majority rule consensus tree is commonly used.

The results for this data are shown in figure 6. Overall, the tree is well resolved except for major uncertainty on one very short internal edge and only moderate support for two deeper internal edges. Checking the mean squared error on the residual resampling replicates after selecting an optimal tree is a good idea, as ideally, it should be very close to that of the original analysis. Since no adjustment was made for the boundary imposed by the non-negativity constraint in the resampling, there will be a bias towards a worse fit upon any *a priori* tree due to this. Acting in the opposite direction, tree selection itself creates a bias towards a better than expected fit by *a posteriori* picking the optimal of a number of different models, trees. The combination of these effects is assessed by comparing the fit of the sum of squares of the original analysis (here, 0.0021705), to the mean of the fit between each replicate data set and its own optimal tree. In this case the mean of 1000 replicates was 0.0021489 with a standard deviation of 0.00042256. Since the difference with the original data (0.0000216) is well within two standard deviations of zero, these biases are judged undetectable and probably inconsequential in this example, suggesting no need to adjust for the effects of tree search and/or the non-negativity constraint.

A bigger issue, as will be shown later, is the effect that Qafzeh6 may be having on the overall estimated variance, which in turn will make the tree appear less stable than it otherwise would be.

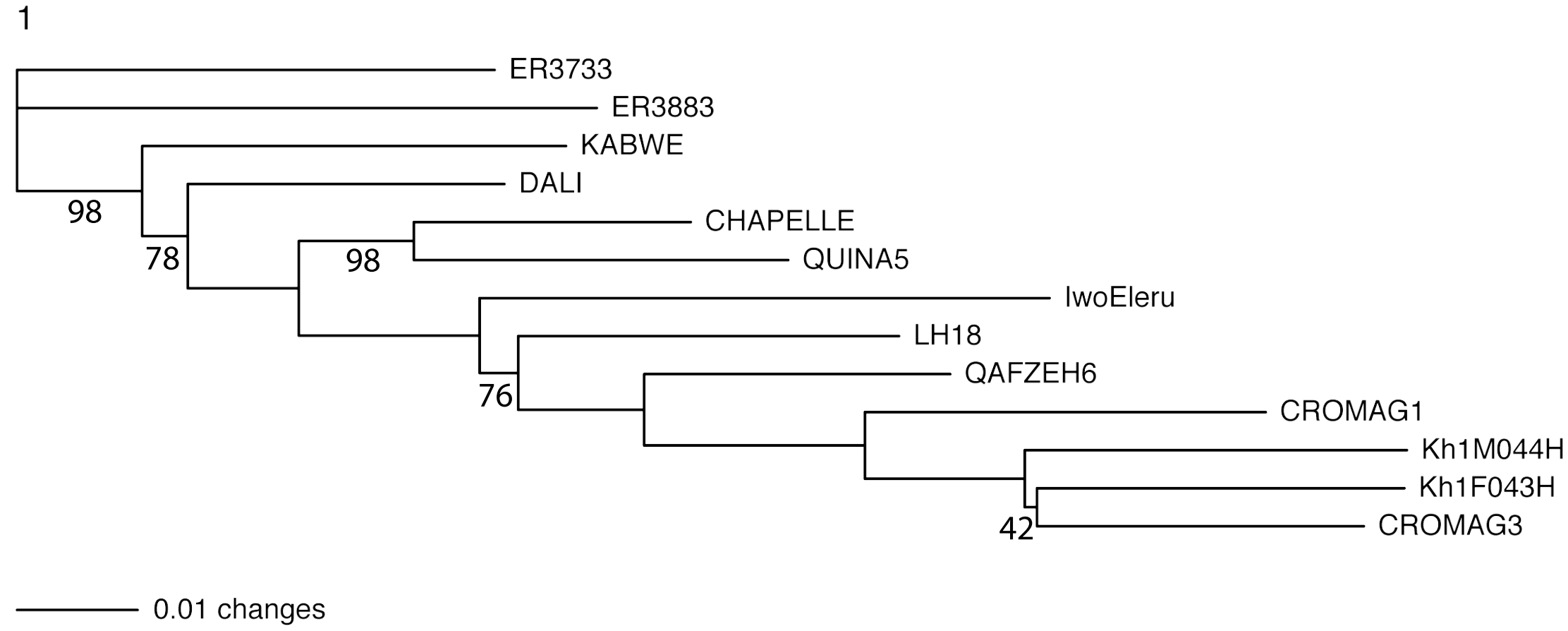

Figure 5. Residual resampling (RR) consensus tree resulting from drawing independent errors $N(0, \hat{\sigma}^2)$, adding them to the expected, tree or patristic distances to create 1000 replicate data sets, and then reanalyzing each using the OLS+ model. Tree redrawn putting the RR percentage, if different to 100, below each edge in the weighted tree.

### 3.5 Checking the residuals

With evidence that OLS+ is the best model on offer for this data, it is appropriate to look at the residuals in detail. There are a number of different ways to look at residuals, each with their own benefits. Various forms of residuals for this data are visualized in figure 6.

Raw Residuals are arrived at by subtracting each entry in the observed distance matrix from the value in the expected (or patristic) distance matrix, that is, $\left(D_{obs_i} - D_{\exp_i}\right)$. They give the user the absolute size of a discrepancy. While the raw residuals are scale dependent, they can be particularly useful to put the discrepancy into a fuller context. For example, with a Hamming distance and genomic data, the size of the residual is a measure of how many derived alleles are involved in producing a particular effect (such as a missing splits' weight, itself perhaps the result of reticulate interbreeding, Waddell et al. 2011b, Waddell 2013). For this data they are shown in figure 6a.

Weighted residuals, which are (observed-expected)/( $w_i \times \sigma^2$ )$^{0.5}$ , where the denominator is the expected standard deviation of that distance, are called standardized residuals. They estimate how many standard errors a particular residual is away from the expectation of zero. The weighted residuals squared are additive and inversely related to the per distance log likelihood function. For visualization, squared residuals may usefully also be signed, but in that form they are no longer additive. Squared standardized residuals are shown in figure 6b.

The sum of squared standardized residuals per taxon in turn can be a gauge of how well that taxon is fitting the model in comparison to other taxa. It needs to be interpreted with some caution, however, since in this context it still carries a weighting implied by the tree structure. For example, five taxa that are all mutually quite badly fitting against all the other taxa, may appear compatible together and may make a sixth taxon that is actually better fitting but uniquely oddly fitting, appear to be the odd one out.

Since the variance is estimated directly from the residuals, and its true scale is unknown, it is important to look at the residuals also with a quantile-quantile, *Q-Q*, plot in order to detect probable outliers. In this case the quantiles are against a *t*-distribution, since the variance is estimated from the data, and the degrees of freedom (d.f.) are equal to the number of informative distances (78) minus fitted parameters (23 edge lengths) minus one for estimating the variance = 54. The *Q-Q* plot for this data is shown in figure 6d.

Considering residuals for this data, a few of the raw residuals appear to be markedly bigger in absolute size than the others (figure 6a). After standardization and squaring the results are shown in figure 6b, which accentuates the suspicion that some of these residuals are abnormally large. The row sums of the squared standardized residuals are, from largest to smallest, QAFZEH6: 13.2, LH18: 8.0, IwoEleru: 5.6, ER3733: 4.6, CHAPELLE: 3.8, Kh1M044H: 3.8, CROMAG1: 3.5, DALI: 2.9, KABWE: 2.4, ER3883: 2.2, Kh1F043H: 2.1, CROMAG3: 1.6 and QUINA5: 1.4. Thus, there is quite a large gap from the taxon with the largest sum of squared residuals to that with the next largest. It therefore appears likely Qafzeh6 is an outlier. Examining the squared residuals in figure 6b, it appears that Qafzeh6 is too unlike L18 and IwoEleru and too like the erectus-like fossil ER3883 in shape to fit onto the tree well. The next largest standardized residual implies that skulls ER3883 and CroMag1 are more convergent on each other than the model expects. Reordering the taxa from left to right according to their sum of squares, the plot of the standardized residuals is shown in figure 6c. Reordering before plotting can be very useful to seeing patterns in the data. Types of reordering that seem useful include by the taxon's sum of squared residuals and by the order the taxa appear in the ordered tree. Finally, 6d shows the *Q-Q* plot and it is seen that the three largest residuals (two positive and one negative) relating to Qafzeh6 are probably outliers, and that it is less clear that the next two largest residuals in absolute size (one positive and one negative) are outliers. Note, the actual size of the residuals on the *Q-Q* plot since everything is scaled by the variance estimated from the same set of squared residuals. What is most important is the deviation from the trend line established by the majority of the residuals and the size of any deviation compared to fluctuations in the trend line. Here the trend line has a slope markedly less than 1 and 1, and the only real fluctuation in it is a smooth dip in the negative residuals before the tail is encountered. The deviations appear much bigger than expected from the dominant trend line and with regard to its own fluctuations. Indeed, with the five biggest outliers removed the data is showing a clear trend towards "short tails."

Thus, in this data, the single taxon that appears to be causing the most misfit is Qafzeh6. The only way to know for sure how much its removal will improve the overall fit is to rerun the whole analysis with it removed. Doing so, the g%SD drops sharply from 6.161 ($k$ = 24) to 4.335 ($k$ = 22) and the tree remains otherwise the same. That is a substantial drop, since the average standardized residual drops by a factor of around 4.335/6.161 or ~30%. The more taxa there are in the analysis, the more profound an average improvement in fit of this order becomes in terms of the average likelihood per piece of data. This means that the treeness of the data has jumped due to the exclusion of just one taxon. The optimal $P$ or $P$' value is now $P$ = 0.077 and again this is not significantly different to $P$ = 0 by a likelihood ratio test.

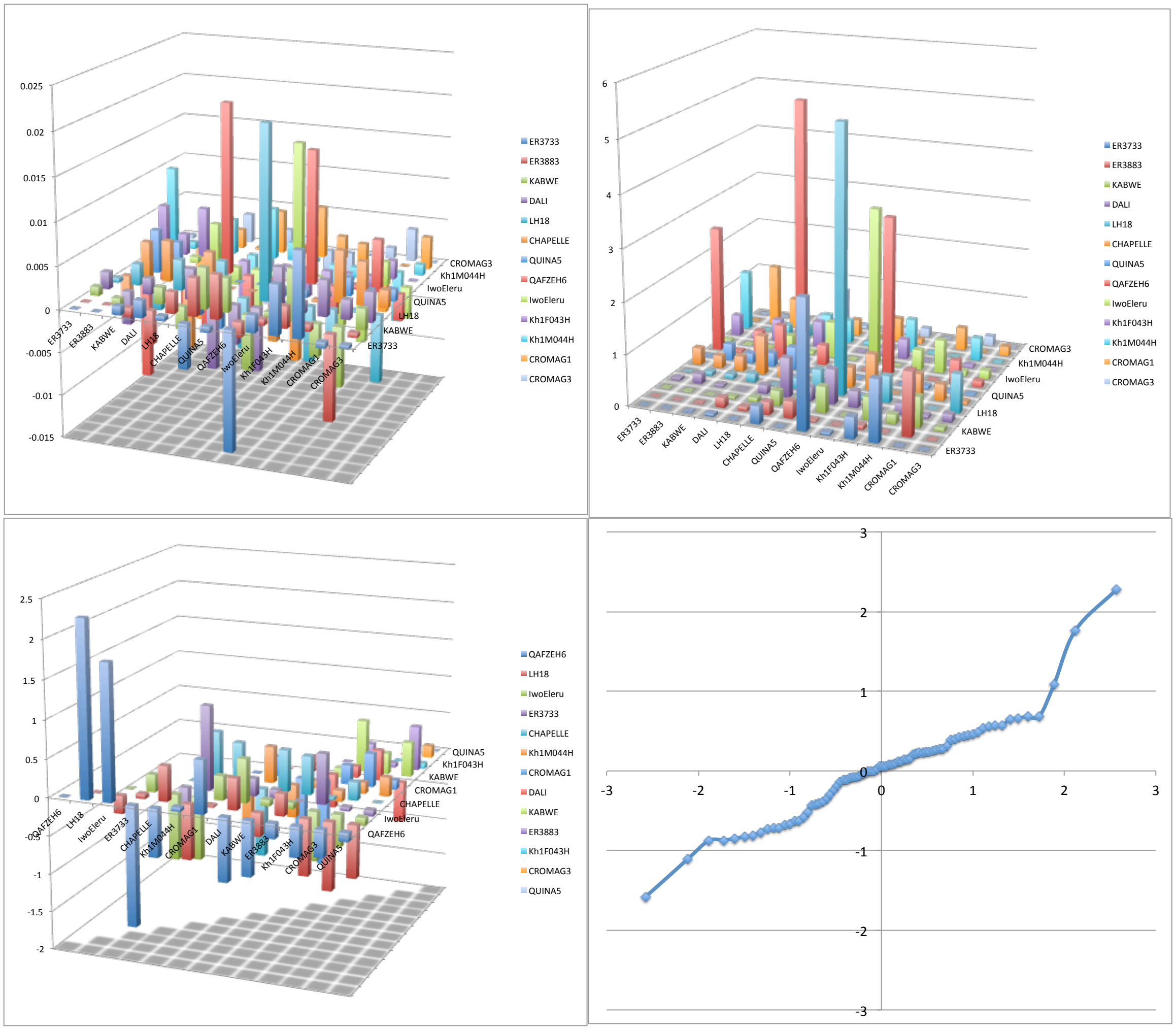

Figure 6. Plotted residuals. (a) Raw residuals. (b) Squared standardized residuals. (c) Standardized residuals with taxa reordered, left to right and forward to back, from most to least deviant. (d) A $Q$-$Q$ plot of expected residuals following a $t$-distribution (degrees of freedom $N - k$ = 78 – 24 = 54) versus observed standardized residuals.

### 3.6 Residual Resampling and NeighborNet

SplitsTree4 is a program that operates with some overlap with PAUP*, but opens up the possibility of fitting planar diagrams as well. There are two main heuristics for inferring a planar graph, NeightborNet (Bryant and Moulton 2004) and Split Decomposition (Bandelt and Dress 1992), with the former being more sensitive and the latter seeming more robust. Planar graphs are a more general type of graph than trees, and can be especially useful in distance-based phylogenetic analyses and for diagnosing non-tree signals as long as the pattern of non or multi-treeness is not too complicated (for example, one reticulate event will result in a distinctive type of graph, being the sum of two trees). A residual resampling script was run, which integrates SplitsTree4 with residual resampling and the g%SD fit statistics. It calls SplitsTree, calculates essential statistics of the fitted planar graph, generates replicate data sets, runs then each replicate to produce its own planar graph, and then summarizes the results.

For the root Procrustes distances, the result of residual resampling applied to the Neighbor next with edge lengths estimated by OLS+ is shown in figure 7. The visualization has been filtered so that any edge with less than 70% resampling support has been removed. The overall g%SD fit has improved markedly from 6.161 ($k$ = 24) to 4.01 ($k$ = 38). Residual resampling support values for splits and edges in figure 7 are generally high, with those related to internal edges of the tree in figure 5 earlier being very similar.

The analyses in figure 7 strongly suggest that Qafzeh6 is an outlier to a phylogenetic tree model. The NeighborNet analysis with residual resampling shows there is strong support for conflicting signals in Qafzeh6. A lot of this is explained if the distance matrix is the sum of two fairly similarly weighted, but different, phylogenetic trees. These two trees could be generated as a consequence of a single reticulate event leading to Qafzeh6 (Waddell 1995, Waddell 2013, with the latter example modeling an example of autosomal gene flow into the Denisovan). One of these two trees has Qafzeh6 sister to the modern humans, while the other one has it moved several internal nodes towards the root, so that it branches just after the Neanderthals diverge from the lineage to moderns. In the NeighborNet the two splits creating the box like structure for Qafzeh6 are of similar size. They are, 0.0171 for the association with modern humans and 0.0107 for the association with Neanderthal and earlier diverging *Homo*. Under an additive, many small genetic effects model, this suggest an ancestry for Qafzeh6 that might be ~62% near modern and 38% archaic. The 98% support value for the split separating Quafzeh6 from all the modern humans suggests that while part of Qafzeh 6's ancestry was near modern, it was not fully modern.

The planar graph of figure 7 thus helps to explain how some of the residuals from the tree might originate. In particular, figure 7 suggests that Qazfeh6 might be a hybrid between a form with a near modern skull and something much more archaic, such as Neanderthals, giving rise to a shape that does not fit well with other pre modern forms from Africa (LH18 and Iwo Eleru). This insight fits well with prior information, with Qazfeh6 being from the Middle East (Palestine, now within Israel), with an age of around 90 kya, during a period when the area was occupied successively by anatomically near modern human populations and populations that appeared to be Neanderthals. It might be expected that in this type of hybridization, if the hybridization was with the Neanderthal-like forms seen in the area, then there should be a split specifically with the two Neanderthals in the sample. Reasons for the split being with deeper archaics generally could be noise in the data, and that the specific signal in this data for Neanderthal calvaria shape is both weak (therefore, little differentiated from an ancestor), and it is the shape of the late classic European Neanderthal upper brain case. It is possible that late Middle Eastern Neanderthals, which are not included in the analysis, may be even less differentiated from an ancestral form in overall shape. It is even possible that Middle Eastern Neanderthal-like forms might even be derived in the direction of modern humans due to near human hybridization into them.

Since these graphs are not directed, the reverse could produce the same effect, that is, that both IwoEleru and LH18 are the hybrids and Qafzeh6 is simply on the lineage leading to modern humans. Against this is the requirement that two individuals, widely separated in time and space show a similar degree and pattern of hybridization. The opportunity for this is also greatly more limited by the geographic locations of these individuals, deep into Africa, where no Neanderthal specimen has ever been reported (Lieberman and Bar-Yosef 2005, Schwartz et al. 2005). That being said a deep reticulate event within Africa for the lineage leading to IwoEleru and LH18, such as interbreeding with an archaic form such as Kabwe, might also explain it. Suggestions that Kabwe might be as young as perhaps 300 kya years ago have been made recently without presenting the data/analyses (Stringer 2012), and have yet to be shown to be robust against other views of a much older estimated age (Millard 2008). The preponderance of the evidence therefore, tends to favor an interaction of near modern humans with Neanderthal-like forms. This is some of clearest evidence yet of the morphological consequences of the hybridization inferred from genomics (Green et al. 2010, Reich et al. 2011, Waddell et al. 2011b). Indeed, while these suggested as little as 3% Neanderthal genes into modern humans, the suggestion here is that during a near modern human Out of Africa event about 90kya, some of the near modern human populations could have had a lot more Neanderthal in them. Whether these Neanderthal genes, or indeed the genes of these near modern humans left descendants in modern humans is less clear. It is possible that this 90kya Out of Africa by near moderns was a double failure in the sense of not populating the rest of the world, nor of contributing significant genes to their near relatives, our ancestors, that went on to overrun the world. The morphology seems to be hinting at this.

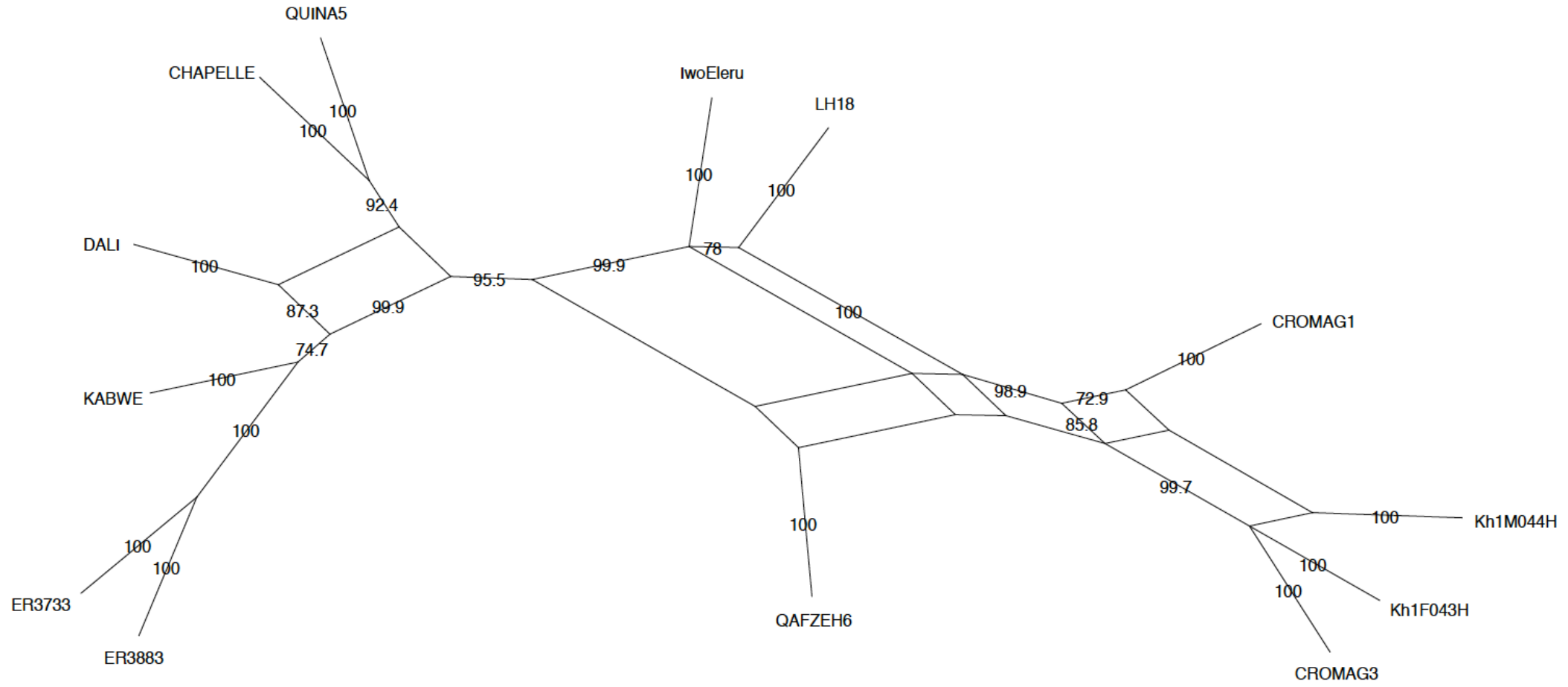

Figure 7. The planar diagram selected by NeighborNet (NN) with edge lengths reconstructed with the OLS+ method. The percentage of the time a split appears in replicates using residual resampling from $N(0, \hat{\sigma}^2)$ is shown, filtered with a cutoff of 0.7. The g%SD is 4.01 ($k$ = 38), which is markedly less than that seen with a tree on this data.

### 3.7 Residual resampling after removing Qafzeh6

The tree with Qafzeh6 removed is shown in figure 8a. The tree resampling results of figure 8a shows considerably higher support values for internal edges. This is consistent with the variance having dropped noticeably. Also shown (figure 8b), is the tree with what is apparently the next most deviant taxon CroMag1, removed. Whether CroMag1 is really an outlier is addressed in detail below, using an array of hopefully more sensitive tests of residuals than those already applied. With CroMag1 removed the residual resampling support for nearly all edges in the tree increases further. Indeed the support for IwoEleru diverging before LH18 is now at 99%, a considerable increases from the 76% seen in the analysis with Qafzeh6 included (figure 5). Interestingly, with Qafzeh6 removed, or this removed plus CroMag1 removed, the optimal BME and NJ tree alters to agree with OLS+ that IwoEleru is more likely to have diverged earlier than LH18, from the lineage leading to modern humans.

Another consequence of removing Qafzeh6 and CroMag1 as an outliers, is that the log likelihood ratio statistic for the test of the monophyly of "Homo heidelbergensis" jumps to 3.515 (g%SD = 3.582, $P$ = 0, $k$ = 20 not monophyletic, versus g%SD = 3.765, $P$ = 0, $k$ = 19 monophyletic). According to the test in Waddell et al. (2002) of the lnL difference of two local rearrangements, this has a p value of 0.0223, which is significant so monophyly is rejected.

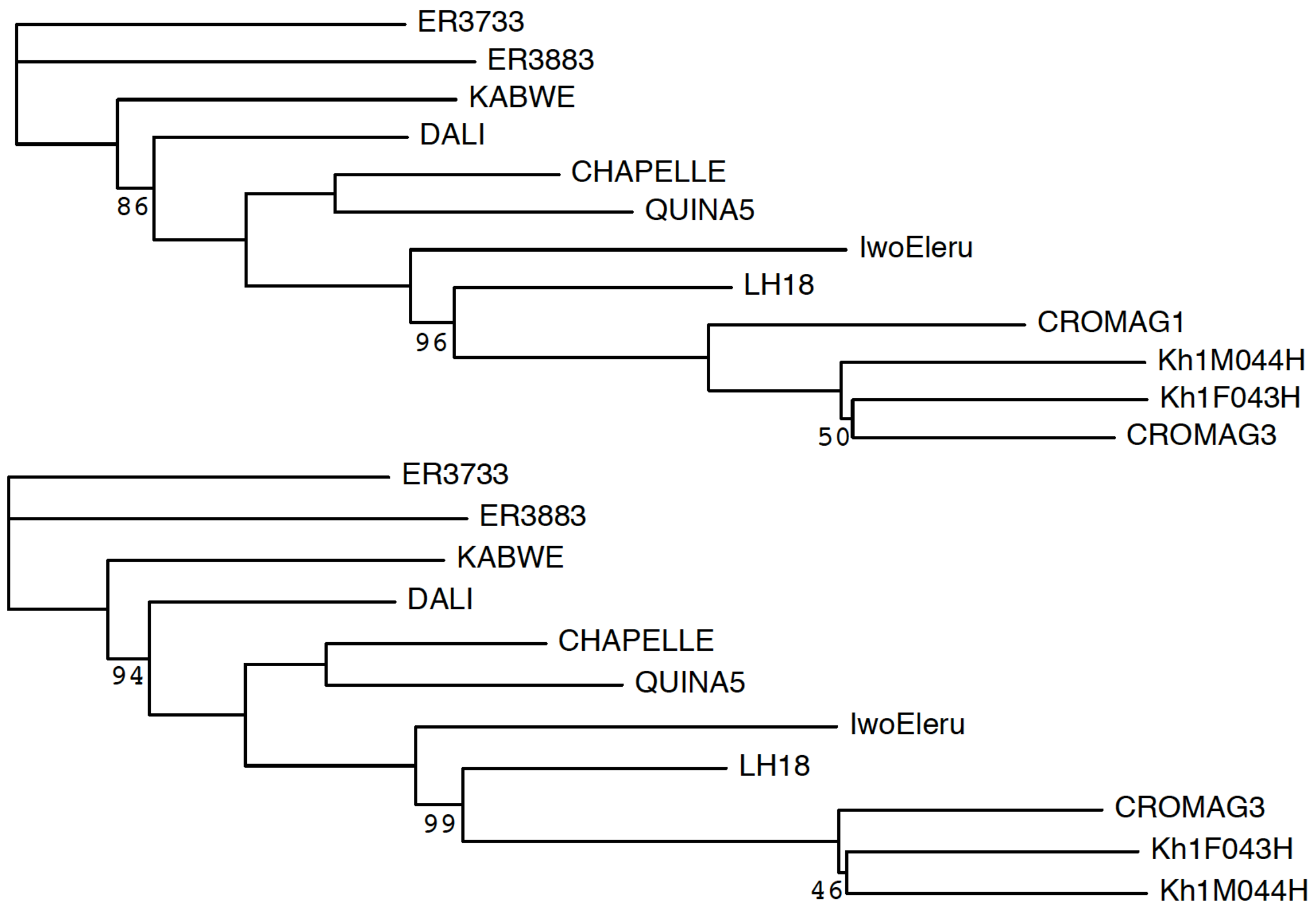

Figure 8. (a) Residual resampling consensus tree OLS+ with residual resampling after Qafzeh5 removed, the g%SD = 4.335 ($k$ = 22) (edges without numbers beside them are at 100%) (b) After CROMAG1removed, the g%SD = 3.724 ($k$ = 20). Figure 8b also involved a slight adjustment inflating the sampling variance by a factor of 1.02 to adjust for a slight shrinkage of the sum of squares due to tree search.

It is also important to check the planar diagram with these same taxa removed. With Qafzeh6 removed, the overall diagram is much more tree-like, as seen in figure 9a. A surprise is that Dali seems to share an unexpected and relatively well-supported (over 90%) secondary relationship with Neanderthals, and Dali is no longer separated by an internal edge from Kabwe. The fit of CroMag1 is also partly modeled as perhaps shared male characteristics with the KhoiSan male.

With CROMAG1 also removed (figure 9b), the Kabwe earlier tree split reappears with around 85% support. Also suggested in the planar diagram is a small convergence of IwoEleru with just the make Neanderthal from Chapelle.

It is useful before proceeding to consider results with the more conservative Split Decomposition method, also available in SplitsTree. With all taxa restored to the analysis, the result is shown in figure 9c. The planar diagram is more tree-like than with NeighborNet and OLS+ edge lengths, but still shows the Qafzeh6 with archaics and Iwo Eleru with Chapelle splits. Removal of Qafzeh6 and repeating the Split Decomposition graph removes all non-tree splits except that grouping IwoEleru and Chapelle (not shown). In this more conservative scenario, Dali is suggested to be a single lineage coming from the same internal node as Kabwe, with no indication of interaction with either earlier or latter archaics.

Clearly, more work is needed on model selection with planar graphs. The residual resampling method certainly suggests feasible archaic interactions when combined with NeighborNet. NeighborNet was developed partly as a response to the impression that Split Decomposition was an overly conservative method. Additionally, the g%SD numbers reported are probably higher than they need to be, with a lot of small, and probably unnecessary, splits included, thus $k$ is much higher than it need be in comparison with the trees or if calculated on only the better supported residual resampling edges.

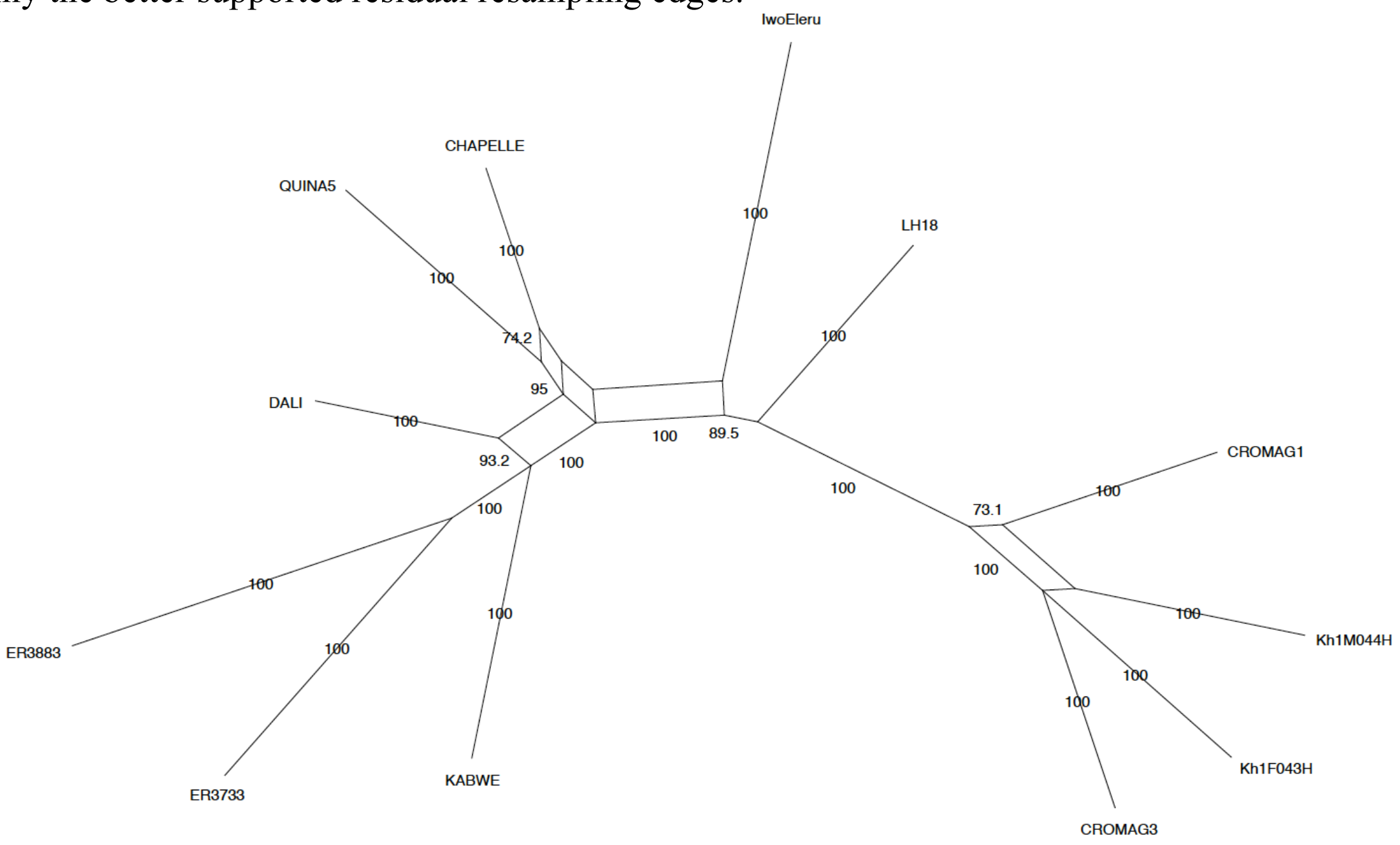

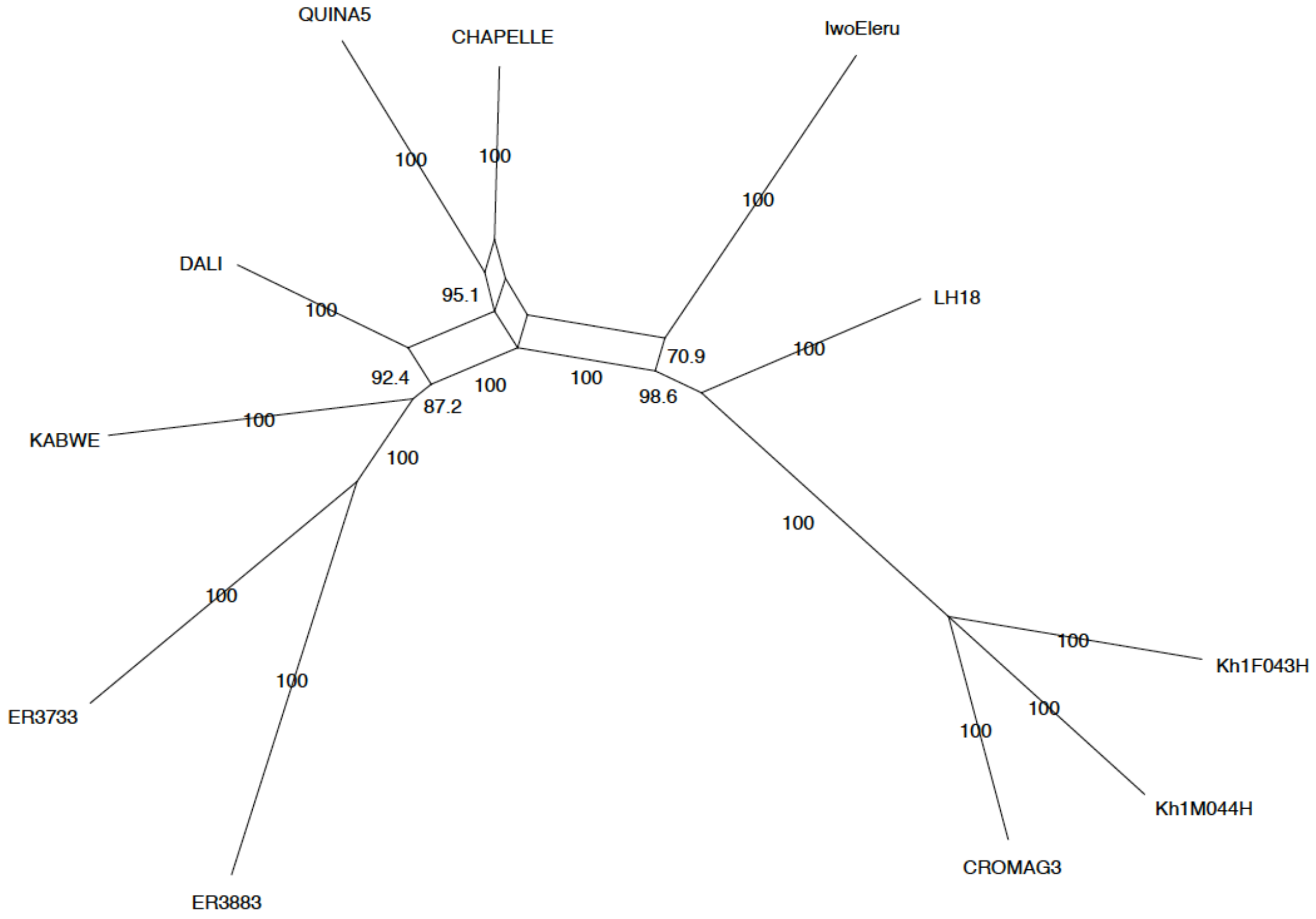

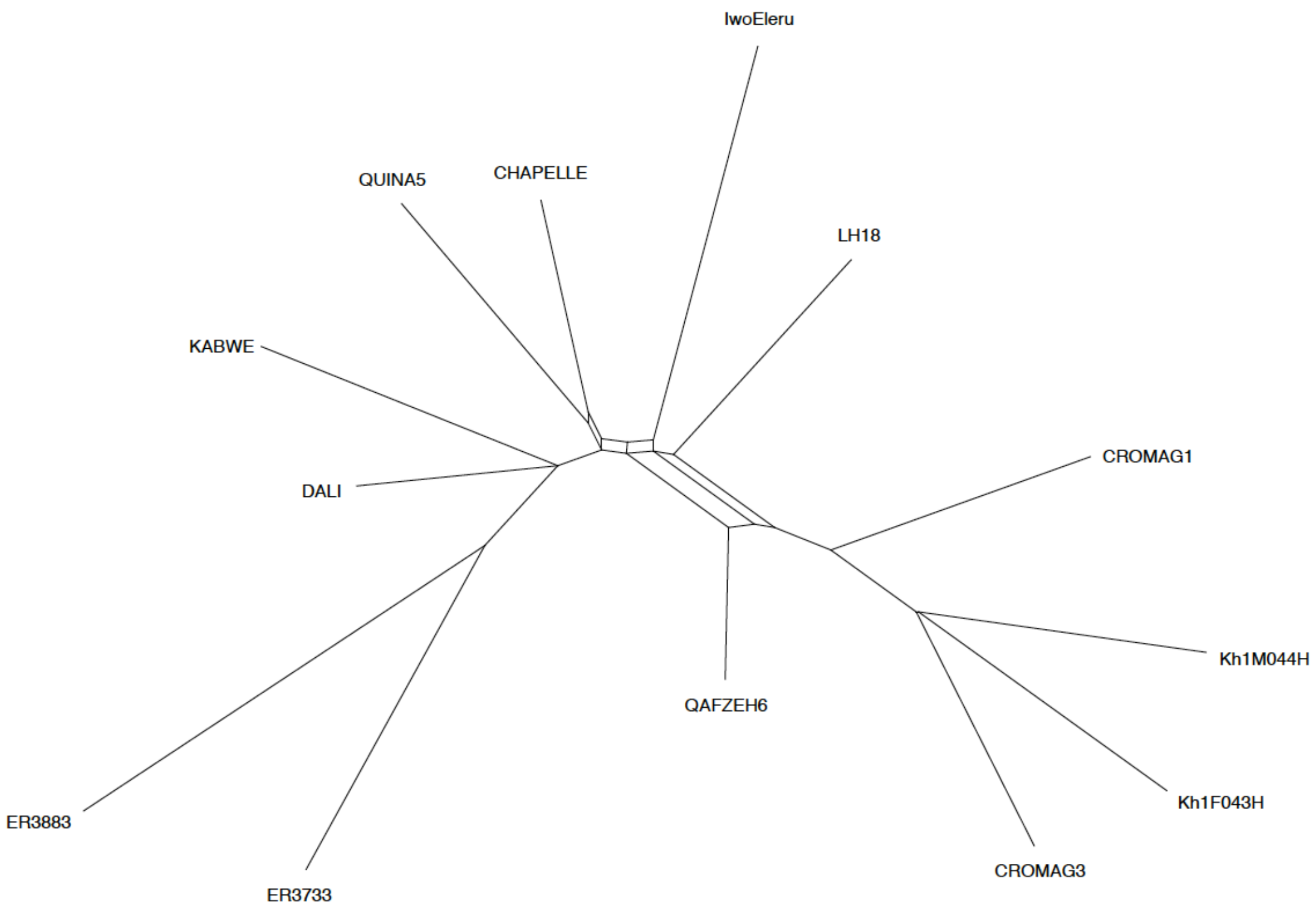

Figure 9. (a) Planar graph selected by NeighborNet (NN) with OLS+ edge lengths and residual resampling after Qafzeh6 is removed. Splits are filtered at a support level of 70% of residual resampling replicates. A variance adjustment of 1.10 was made due to graph search shrinking the sum of squares. g%SD = 3.341 ($k$ = 33) (b) After CroMag1 is removed and filtered at a support level of 70%, g%SD = 3.391 ($k$ = 29). (c) The Split Decomposition graph for these taxa, a more conservative method than NN that estimates edge lengths with its own formulae, not OLS+.

### 3.8 The covariance matrix of edge lengths

It is useful at this point to list a set of closed form solutions for edge length estimates and other useful quantities, before going on to look at the structure of edge length estimates on the optimal OLS+ tree. These formulae are give estimates assuming no boundaries in the parameter space (such as, edge lengths can be negative). The distance data in **D** will be re-expressed as a column vector, **d**, where the columns of the lower triangle of **D**, less the diagonal, are concatenated. Let the design matrix, **X**, have $N$ rows corresponding to the $t(t$-1)/2 informative distances in the same order as they appear in **d**, and let each column correspond to an edge in the tree under consideration. In this design matrix there is 1 if an edge is crossed by a certain distance, 0 otherwise. We will use $\mathbf{d}_{obs}$ to indicate the observed distances and $\mathbf{d}_{exp}$ for the expected distances of the model (tree). Further, let the optimal edge lengths be contained in a column vector **b**, with entries appearing in the same order as those of the columns of **X**. Then $\mathbf{b} = (\mathbf{X}^t\mathbf{X})^{-1}\mathbf{X}^t\mathbf{d}_{obs}$, while $\mathbf{d}_{exp} = \mathbf{Xb}$, and residual sum of squares is given by SS = $(\mathbf{d}_{obs}\text{-}\mathbf{d}_{exp})^t(\mathbf{d}_{obs}\text{-}\mathbf{d}_{exp})$. Let **V** be the variance-covariance matrix of the distances, which is a diagonal matrix with non-zero entries equal to SS/$N$ = $\sigma^2$. The variance covariance matrix of **b** is obtained as $\mathbf{V_b} = (\mathbf{X}^t\mathbf{V}^{-1}\mathbf{X})^{-1}$, which simplifies to $\mathbf{V_b} = (\mathbf{X}^t\mathbf{X})^{-1}\sigma^2$, for ordinary least squares. These formulae are standard results for linear models (e.g., Bulmer 1989). A lot of these matrix multiplications can be sped up tremendously using techniques such as those described in Bryant and Waddell (1998).

Leverage effects, which are one distance being able, or forced, to fit the model more closely than others, can be estimated in linear models via the hat matrix, $\mathbf{H} = \mathbf{X}(\mathbf{X}^t\mathbf{V}^{-1}\mathbf{X})^{-1}\mathbf{X}^t\mathbf{V}^{-1}$, which simplifies to $\mathbf{H} = \mathbf{X}(\mathbf{X}^t\mathbf{X})^{-1}\mathbf{X}^t$ with OLS. The diagonal elements of this matrix are the leverage of each distance on the tree defined by **X**. Studentized residuals that are further adjusted by the factor 1/(1-leverage)$^2$ are called studentized residuals. It is useful to note that with distance methods, a distance that on the tree being considered is between the two tips of a cherry, will fit perfectly (except in extraordinary circumstances such as violation of the triangle inequality), and hence it has a leverage of 1 and a residual of zero.

The variance-covariance matrix of the parameter estimates is given by the equation $\mathbf{V_b} = (\mathbf{X}^t\mathbf{V}^{-1}\mathbf{X})^{-1}$, which simplifies to $\mathbf{V_b} = (\mathbf{X}^t\mathbf{X})^{-1}\sigma^2$, where $\sigma^2$ is a scalar equal to the mean squared error or SS/*N*. The structure of the variance-covariance matrix for this type of model is shown in figure 10 with the order of edges matching the tree shown in figure 1. Note that under the OLS tree model an edge length shows appreciable correlation, positive or negative, only with that of the other edges it is adjacent to. Edge lengths of the tips of a cherry are positively correlated with each other and negatively correlated with the length of the edge leading to them. This is a consequence of their combined length being the distance between these two tips, and when this distance is larger than expected, their sum will increase at the expense of the length of the edge leading directly to them. A weaker set of correlations of the same form involves a terminal edge not in a cherry being positively correlated with the internal edge towards the outside of the tree it is connected to and negatively correlated with the more internal edge it is connected to. The deeper into the tree, the weaker this effect becomes.

| | K3733 | K3883 | 1 | KABWE | 2 | DALI | 3 | 4 | CHAPELLE | QUINA5 | 5 | IwoEleru | 6 | LH18 | 7 | CROMAG1 | 8 | Kh1M044H | 9 | Kh1F043H | CROMAG3 |
|---|---|---|---|---|---|---|---|---|---|---|---|---|---|---|---|---|---|---|---|---|---|
| K3733 | 1.0 | 0.7 | -0.7 | 0.0 | 0.0 | 0.0 | 0.0 | 0.0 | 0.0 | 0.0 | 0.0 | 0.0 | 0.0 | 0.0 | 0.0 | 0.0 | 0.0 | 0.0 | 0.0 | 0.0 | 0.0 |
| K3883 | 0.7 | 1.0 | -0.7 | 0.0 | 0.0 | 0.0 | 0.0 | 0.0 | 0.0 | 0.0 | 0.0 | 0.0 | 0.0 | 0.0 | 0.0 | 0.0 | 0.0 | 0.0 | 0.0 | 0.0 | 0.0 |
| 1 | -0.7 | -0.7 | 1.0 | 0.3 | -0.4 | 0.0 | 0.0 | 0.0 | 0.0 | 0.0 | 0.0 | 0.0 | 0.0 | 0.0 | 0.0 | 0.0 | 0.0 | 0.0 | 0.0 | 0.0 | 0.0 |
| KABWE | 0.0 | 0.0 | 0.3 | 1.0 | -0.5 | 0.0 | 0.0 | 0.0 | 0.0 | 0.0 | 0.0 | 0.0 | 0.0 | 0.0 | 0.0 | 0.0 | 0.0 | 0.0 | 0.0 | 0.0 | 0.0 |
| 2 | 0.0 | 0.0 | -0.4 | -0.5 | 1.0 | 0.2 | -0.5 | 0.0 | 0.0 | 0.0 | 0.0 | 0.0 | 0.0 | 0.0 | 0.0 | 0.0 | 0.0 | 0.0 | 0.0 | 0.0 | 0.0 |
| DALI | 0.0 | 0.0 | 0.0 | 0.0 | 0.2 | 1.0 | -0.4 | 0.0 | 0.0 | 0.0 | 0.0 | 0.0 | 0.0 | 0.0 | 0.0 | 0.0 | 0.0 | 0.0 | 0.0 | 0.0 | 0.0 |
| 3 | 0.0 | 0.0 | 0.0 | 0.0 | -0.5 | -0.4 | 1.0 | 0.0 | 0.0 | 0.0 | -0.3 | 0.0 | 0.0 | 0.0 | 0.0 | 0.0 | 0.0 | 0.0 | 0.0 | 0.0 | 0.0 |
| 4 | 0.0 | 0.0 | 0.0 | 0.0 | 0.0 | 0.0 | 0.0 | 1.0 | -0.8 | -0.8 | -0.1 | 0.0 | 0.0 | 0.0 | 0.0 | 0.0 | 0.0 | 0.0 | 0.0 | 0.0 | 0.0 |
| CHAPELLE | 0.0 | 0.0 | 0.0 | 0.0 | 0.0 | 0.0 | 0.0 | -0.8 | 1.0 | 0.7 | 0.0 | 0.0 | 0.0 | 0.0 | 0.0 | 0.0 | 0.0 | 0.0 | 0.0 | 0.0 | 0.0 |
| QUINA5 | 0.0 | 0.0 | 0.0 | 0.0 | 0.0 | 0.0 | 0.0 | -0.8 | 0.7 | 1.0 | 0.0 | 0.0 | 0.0 | 0.0 | 0.0 | 0.0 | 0.0 | 0.0 | 0.0 | 0.0 | 0.0 |
| 5 | 0.0 | 0.0 | 0.0 | 0.0 | 0.0 | 0.0 | -0.3 | -0.1 | 0.0 | 0.0 | 1.0 | -0.1 | -0.5 | 0.0 | 0.0 | 0.0 | 0.0 | 0.0 | 0.0 | 0.0 | 0.0 |
| IwoEleru | 0.0 | 0.0 | 0.0 | 0.0 | 0.0 | 0.0 | 0.0 | 0.0 | 0.0 | 0.0 | -0.1 | 1.0 | 0.0 | 0.0 | 0.0 | 0.0 | 0.0 | 0.0 | 0.0 | 0.0 | 0.0 |
| 6 | 0.0 | 0.0 | 0.0 | 0.0 | 0.0 | 0.0 | 0.0 | 0.0 | 0.0 | 0.0 | -0.5 | 0.0 | 1.0 | -0.3 | -0.5 | 0.0 | 0.0 | 0.0 | 0.0 | 0.0 | 0.0 |
| LH18 | 0.0 | 0.0 | 0.0 | 0.0 | 0.0 | 0.0 | 0.0 | 0.0 | 0.0 | 0.0 | 0.0 | 0.0 | -0.3 | 1.0 | 0.1 | 0.0 | 0.0 | 0.0 | 0.0 | 0.0 | 0.0 |
| 7 | 0.0 | 0.0 | 0.0 | 0.0 | 0.0 | 0.0 | 0.0 | 0.0 | 0.0 | 0.0 | 0.0 | 0.0 | -0.5 | 0.1 | 1.0 | -0.4 | -0.5 | 0.0 | 0.0 | 0.0 | 0.0 |
| CROMAG1 | 0.0 | 0.0 | 0.0 | 0.0 | 0.0 | 0.0 | 0.0 | 0.0 | 0.0 | 0.0 | 0.0 | 0.0 | 0.0 | 0.0 | -0.4 | 1.0 | 0.2 | 0.0 | 0.0 | 0.0 | 0.0 |
| 8 | 0.0 | 0.0 | 0.0 | 0.0 | 0.0 | 0.0 | 0.0 | 0.0 | 0.0 | 0.0 | 0.0 | 0.0 | 0.0 | 0.0 | -0.5 | 0.2 | 1.0 | -0.5 | -0.4 | 0.0 | 0.0 |
| Kh1M044H | 0.0 | 0.0 | 0.0 | 0.0 | 0.0 | 0.0 | 0.0 | 0.0 | 0.0 | 0.0 | 0.0 | 0.0 | 0.0 | 0.0 | 0.0 | 0.0 | -0.5 | 1.0 | 0.3 | 0.0 | 0.0 |
| 9 | 0.0 | 0.0 | 0.0 | 0.0 | 0.0 | 0.0 | 0.0 | 0.0 | 0.0 | 0.0 | 0.0 | 0.0 | 0.0 | 0.0 | 0.0 | 0.0 | -0.4 | 0.3 | 1.0 | -0.7 | -0.7 |
| Kh1F043H | 0.0 | 0.0 | 0.0 | 0.0 | 0.0 | 0.0 | 0.0 | 0.0 | 0.0 | 0.0 | 0.0 | 0.0 | 0.0 | 0.0 | 0.0 | 0.0 | 0.0 | 0.0 | -0.7 | 1.0 | 0.7 |
| CROMAG3 | 0.0 | 0.0 | 0.0 | 0.0 | 0.0 | 0.0 | 0.0 | 0.0 | 0.0 | 0.0 | 0.0 | 0.0 | 0.0 | 0.0 | 0.0 | 0.0 | 0.0 | 0.0 | -0.7 | 0.7 | 1.0 |

Figure 10. A heat map of the matrix of correlations of edge lengths. The order of edges is that which appear in figure 1 reading down the page. The dominant structure is that an edge has appreciable correlation only with the edges it shares a node with. This very localized correlation structure is a property of many exponentially weighted models where the weight is on $\mathbf{D}_{exp}$, and it is what allows very fast tree search algorithms to be devised (e.g. fastME, Desper and Gascuel 2002).

If the variance is estimated from the data, then it to is treated as a parameter when it is used to further estimate quantities such as the likelihood of the data. It's variance is $2\sigma^4/(N\text{-}k)$ in a reduced bias form or $2\sigma^4/N$ in its maximum likelihood form. Since the variance is orthogonal to the edge length parameters, its covariance with all of them is zero.

The standard error of parameters is derived from the diagonal of the variance-covariance matrix. Let us now examine the edge length, its standard error (s.e.) and its length/s.e. (underlines) with edges appearing in the order they appear down the page in figure 1. That is, K3733: 0.0403, 0.0025, <u>15.9</u>; K3883: 0.0475, 0.0025, <u>18.8</u>; 1: 0.0104, 0.0030, <u>3.5</u>; KABWE:

0.0351, 0.0019, 18.6; **2: 0.0038, 0.0023, 1.6;** DALI: 0.0262, 0.0016, 16.1; 3: 0.0096, 0.0018, 5.3; 4: 0.0093, 0.0026, 3.6; CHAPELLE: 0.0232, 0.0025, 9.2; QUINA5: 0.0308, 0.0025, 12.2; 5: 0.0170, 0.0017, 10.0; IwoEleru: 0.0452, 0.0015, 30.9; **6: 0.0045, 0.0020, 2.3;** LH18: 0.0287, 0.0015, 19.0; 7: 0.0263, 0.0021, 12.7; CROMAG1; 0.0327, 0.0016, 20.0; 8: 0.0137, 0.0023, 5.9; Kh1M044H: 0.0315, 0.0019, 16.7; **9: 0.0012, 0.0030, 0.4;** Kh1F043H:0.0305, 0.0025, 12.0; CROMAG3:0.0271, 0.0025, 10.7. The three weakest supported edges are shown in bold. These are also the three edges of the tree that have weakest residual resampling support. The edge separating IwoEleru from LH18 is the third weakest, but is over two standard errors from zero, so, in its own right fairly significant (this does not take into account that with residual resampling it has direct competition from two other edges that can replace it without altering the rest of the tree). The length/s.e. values are *t*-statistics since the variance is estimated from a sample, and since the degrees of freedom is > 50, they are reasonably safely treated as *z*-statistics.

It is interesting to note that the CroMag1 skull is separated by a deep internal edge of over 5 standard errors in length. This adds to the growing impression that CroMag1 is not a typical modern human in shape.

### 3.9 Tree assessment via a parsimonious model

Was it, or was it not, William of Occam that suggested the best explanation is that which explains the situation with the fewest additional assumptions? More recently, this principal has increasingly been associated with information theory, which has arisen with the development of electronic signals and computers. In its purist form, Occam's suggestion today might be equated with Algorithmic Information Criteria, such as the shortest possible program to reproduce the data. Unfortunately, these are incomputable, but two popular information criteria for parsimonious model selection are AIC and BIC. AIC rests on the assumption that the best model is that which would most accurately predict another sample from the unknown true model (Akaike 1974). Under some assumptions that are fairly general and arguably robust, although rarely exactly met, minimizing the criterion $\text{AIC} = -2\ln L - 2k$ will minimize the expected Kullback-Leibler distance to the new sample. For small $N$ and/or large $k$, a corrected statistic called AICc is important. It is of the form $\text{AICc} = -2\ln L\text{-}2kN/(N\text{-}k)$ (Sugiura 1978).

BIC, in contrast, has a different philosophical basis, and its aim to identify the model that is expected to yield the highest posterior probability as more data are added (Schwartz 1978). To do this, it discards the effect of priors and concentrates on the dominant term in the posterior probability of a model, as the amount of data increases. This is done with a Lagrange approximation of the marginal likelihood under the assumption of normal errors and no boundaries, and yields $\text{BIC} = -2\ln L\text{-}k\ln(N)$.

Interestingly, there are relatively simple modifications of g%SD that are inversely monotonic with AIC, AICc and BIC rather than with the $\ln L$ (Waddell and Tan 2012). For g%AIC replace the $1/(N - k)$ multiplier of the sum of squares with $\exp(2k/N)/N$. For AICc, replace the term $1/(N - k)$ with the term $\exp(2k/(N\text{-}k))/N$. For g%BIC replace the term $1/(N - k)$ with the term $N^{(k/N-1)}$.

Due to the simple covariance structure of OLS, and the fact that most of the edges which are likely to be dropped from consideration by these criteria are separated by at least one strongly supported internal edge, the simple order of reverse selection with respect to the length of an edge in standardized units, that is, **9: 0.4; 2: 1.6; 6: 2.3;** 1: 3.5; 4: 3.6; 3: 5.3; 8: 5.9; 5: 10.0; 7: 12.7**;** is fairly likely to identify the optimal set of edges to remove in each case. The results are shown in table 1. For AIC, the optimal tree is the optimal binary tree with the shortest internal edge collapsed, for AICc it is this tree with the internal edge between the "*Homo heidelbergensis*" skulls also removed, and for BIC it is again this tree. The g%AICc statistic of 5.641 might be an indication the g%SD expected if this model with its parameter estimates could be applied to another equivalent data set from the same true model or process. Curiously, none of these model selection criteria elected to collapse the internal edge separating Iwo Eleru from LH18.

This concludes the examination of model selection criteria, except to mention that there are other alternative philosophies in information theory that may be applied and lead to their own formulae for decisions. One of the most interesting of these are the stochastic complexity criteria, including normalized maximum likelihood (Rissanen 2007), which are an attempt to apply the "Kolmogorov Criterion" with additional assumptions and critically, be computable.

Table 1. Parsimonious selection of edges in the OLS+ tree of 12 taxa by backwards selection on the optimal OLS+ tree using g%AIC, g%AICc and g%BIC. SS are multiplied by $10^8$. The optimal tree identified by each criterion is highlighted in bold.

| Edge deleted | *k* | SS | g%SD | g%AIC | g%AICc | g%BIC |
|---|---|---|---|---|---|---|
| none | 22 | 9132 | 4.335 | 4.940 | 5.836 | 7.116 |
| 9 | 21 | 9168 | **4.295** | **4.875** | 5.656 | 6.907 |
| 2 | 20 | 9722 | 4.375 | 4.945 | **5.641** | **6.890** |
| 6 | 19 | 10878 | 4.578 | 5.152 | 5.788 | 7.061 |
| 1 | 18 | 15474 | 5.403 | 6.052 | 6.704 | 8.158 |
| 4 | 17 | 18187 | 5.797 | 6.463 | 7.067 | 8.568 |

### 3.10 More intense probing of residuals

Above, with 13 taxa, we went over some simple to calculate ways of detecting residuals, and an outlier was apparent. We now consider the fit of the 12 taxa distance data to the OLS+ tree using more rigorous and hopefully sensitive methods.

Figure 11 shows the results for four different types of residual, in order of increasing complexity. These are standardized residuals, standardized residuals with the variance estimated "externally" by excluding that residual, internally studentized residuals and externally studentized residuals. The first method might therefore more formally be called internally standardized residuals. Also, since externalization involves leaving out one observation at a time, it is also sometimes referred to as a kind of jackknifing of the variance.

The studentized methods weight the standardized residuals by multiplying them by $1/\sqrt{1-\hat{h}_i^2}$ , where $\hat{h}_i^2$ is the estimated leverage of the *i*-th distance estimated from the so-called hat matrix described above. The standardized residuals plotted previously were treated as though the true variance is estimated from a very large number of residuals and they are assumed to be normally distributed. However, since the degrees of freedom in estimating the variance has now dropped to less than 50, their true *t*-distributed character should be taken into account. For the *Q-Q* plot in figure 11a, the internally studentized residuals of the data (*y*-axis) are plotted against their expected values, which are $sign\sqrt{B(1/2,(N-k-1)/2)}\times(N-k-1)$, that is the signed square root of a beta distribution multiplied by the residual degrees of freedom (Stuart and Ord 1987, 1990). The externally studentized residuals have a simpler expected distribution, that is a Student's *t*-distribution with degrees of freedom set to the residual degrees of freedom minus one, here 66 - 21-1 = 44.

Figure 11 suggests that, if anything, the standardized residual plot is slightly S shaped. Externalization makes little difference here, although in the earlier example with pronounced outliers, it would have accentuated the largest outliers further. The application of studentization here sees the residuals fit even more closely to expectations, and the slight S-shaped distribution is straightened slightly. However, since neither internal nor external studentization makes a clearly perceptible difference, it further reinforces the suggestion of no apparent outliers.

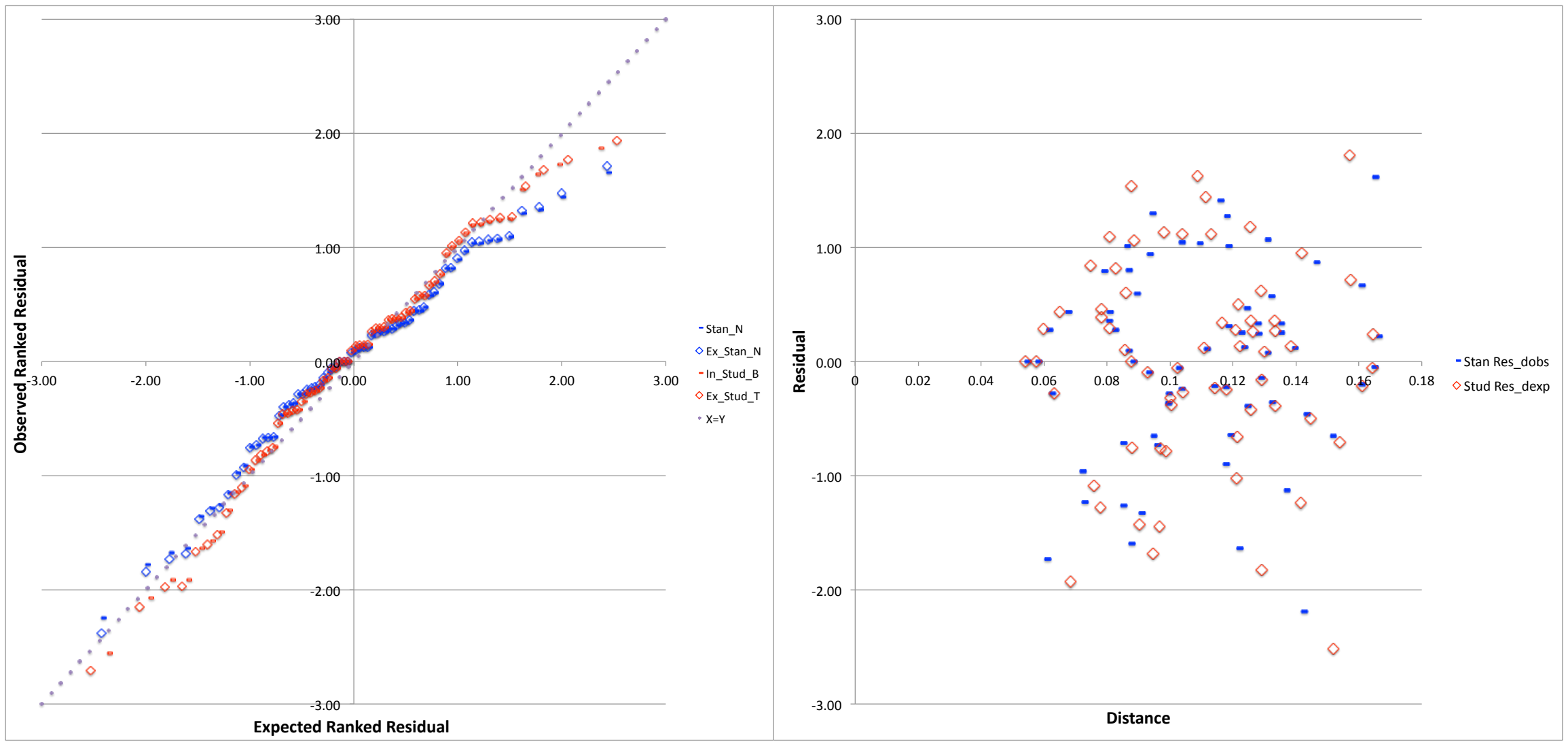

Figure 11. (a) A *Q-Q* plot for standardized and studentized residuals. (b) Plot of internally standardized residuals versus the observed distance and externally studentized residuals versus expected distances.

Figure 11b is a graphical check to see if there is any trend to the residuals. No clear trend of size of the residual with respect to the size of the distance is seen, consistent with the test of $P = 0$ failing to reject, with the actual power $P$ being barely positive. The largest negative residual does, however, stand out a little. There is also a suggestion of some web-like clumpedness in the residuals, caused mostly by a North East to South West trending void for distances of ~0.06 to 0.13. It is hard to quantify such patterns as truly non-random. The simple standardized residuals suggest a very similar result to the externally studentized residuals in this case.

While there do not seem to be any clear outliers or trends thus far, there are a further examinations that might be useful. The first is a calculation and plot of Cook's distances. The Cook's distance (CD) for the *i*-th observed distance is equal to $CD_i = \sum_{j=1}^{N} (D_{\exp_j} - D_{\exp_{j(i)}})^2 / (kMSE)$, that is, the sum of squares over all the pairwise expected (model) distances calculated from edge lengths of the tree with that observed distance excluded, divided by the mean square error multiplied by $k$. An alternative expression is, $CD_i = e_i^2 / (kMSE) \times [h_{ii} / (1 - h_{ii})]$, where $e_i^2$ is the raw residual squared and $h_{ii}$ is the leverage obtained via the hat matrix (Stuart and Ord 1987, 1990). From this we see that $CD_i$ is a akin to a standardized residual, with an extra weighting for its leverage. As such, its value may usefully be shown in a bubble plot with axes for the studentized residual and leverage and the size of the bubble being proportional to the Cook's distance, as seen in figure 12a.

Figure 12a suggests that the Cook's distances in general are reflecting the size of the externally studentized residual. Further, there are no particularly atypical residuals, which have large Cook's distances due mostly to having large leverages. Figure 13a shows a combined table/heat map of the Cooks distances for each distance. There are a number of Cook's distances larger than 1. What seems to be more telling is that a large number of them appear to be associated with CroMag1. A very similar pattern is reflected in the squared studentized residuals of figure 14a. Figure 12b suggests that only one of the residuals of CroMag1 plots to a markedly bigger Cook's distance than expected and this is CroMag1 to the KhoiSan Male. It has a negative sign, and might conceivably be due to some convergence in shape due to maleness (it is widely assumed by paleontologists that CroMag1 is a male, while CroMag3 is largely treated as indeterminate sex or female, e.g. Schwartz et al. 2005). Figure 12c shows that there is a strong relationship between the sum of Cook's distances and the sum of squared studentized residuals for each taxon, corroborating the usefulness of this easy to calculate per taxon measure. The convergence of CroMag1 and the KhoiSan male is a feature also picked up in the earlier analyses using NeighborNet.

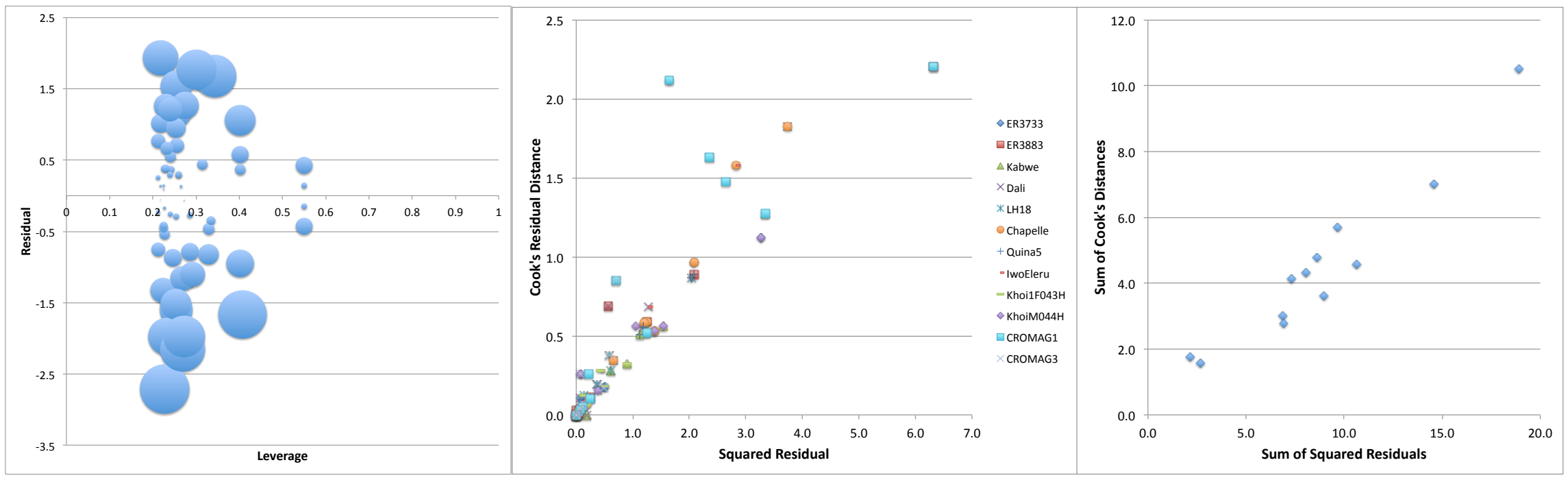

Figure 12. (a) A plot of residual (externally studentized) magnitude on the *y*-axis against its leverage with the size of the bubble being proportional to Cook's distance, on the *x*-axis. (b) A plot of the Cook's distance against the squared externally studentized residual. (c) A plot of the sum of sum of Cook's distance for each taxon versus the sum of squared externally studentized residuals for that taxon. The order of the taxa on the *x*-axis follows that of the last column in figure 14b.

Figure 13 considers the sums of various divergence measures. Figure 13a has the sum of Cooks distances per taxon. The sum of squares per taxon is assumed to be approximately the sum of 11 uncorrelated residuals, which are approximated as a chi-square with 11 degrees of freedom. A *Q-Q* plot of these row sums is shown in figure 13b. There is a hint that taxa CroMag1 and Chapelle are outliers while CroMag3 and the KhoiSan female appear almost too well fitting.

| | | | | | | | | | | | | | |
|---|---|---|---|---|---|---|---|---|---|---|---|---|---|
| CROMAG1 | 0.00 | 0.52 | 2.12 | 1.48 | 2.21 | 0.05 | 1.63 | 1.27 | 0.03 | 0.10 | 0.85 | 0.26 | **10.51** |
| CHAPELLE | 0.52 | 0.00 | 0.53 | 1.58 | 0.59 | 1.82 | 0.35 | 0.58 | 0.00 | 0.97 | 0.01 | 0.07 | 7.02 |
| Khoi1M044H | 2.12 | 0.53 | 0.00 | 0.56 | 0.00 | 0.15 | 0.07 | 0.57 | 0.05 | 1.12 | 0.26 | 0.26 | 5.69 |
| IwoEleru | 1.48 | 1.58 | 0.56 | 0.00 | 0.04 | 0.68 | 0.10 | 0.01 | 0.00 | 0.03 | 0.03 | 0.28 | 4.80 |
| ER3883 | 2.21 | 0.59 | 0.00 | 0.04 | 0.00 | 0.69 | 0.11 | 0.03 | 0.89 | 0.00 | 0.02 | 0.02 | 4.58 |
| Dali | 0.05 | 1.82 | 0.15 | 0.68 | 0.69 | 0.00 | 0.19 | 0.00 | 0.56 | 0.11 | 0.05 | 0.01 | 4.31 |
| LH18 | 1.63 | 0.35 | 0.07 | 0.10 | 0.11 | 0.19 | 0.00 | 0.28 | 0.87 | 0.02 | 0.38 | 0.12 | 4.12 |
| Kabwe | 1.27 | 0.58 | 0.57 | 0.01 | 0.03 | 0.00 | 0.28 | 0.00 | 0.51 | 0.03 | 0.01 | 0.33 | 3.61 |
| QUINA5 | 0.03 | 0.00 | 0.05 | 0.00 | 0.89 | 0.56 | 0.87 | 0.51 | 0.00 | 0.03 | 0.00 | 0.06 | 3.01 |
| ER3733 | 0.10 | 0.97 | 1.12 | 0.03 | 0.00 | 0.11 | 0.02 | 0.03 | 0.03 | 0.00 | 0.17 | 0.18 | 2.77 |
| CROMAG3 | 0.85 | 0.01 | 0.26 | 0.03 | 0.02 | 0.05 | 0.38 | 0.01 | 0.00 | 0.17 | 0.00 | 0.00 | 1.77 |
| Khoi1F043H | 0.26 | 0.07 | 0.26 | 0.28 | 0.02 | 0.01 | 0.12 | 0.33 | 0.06 | 0.18 | 0.00 | 0.00 | 1.58 |

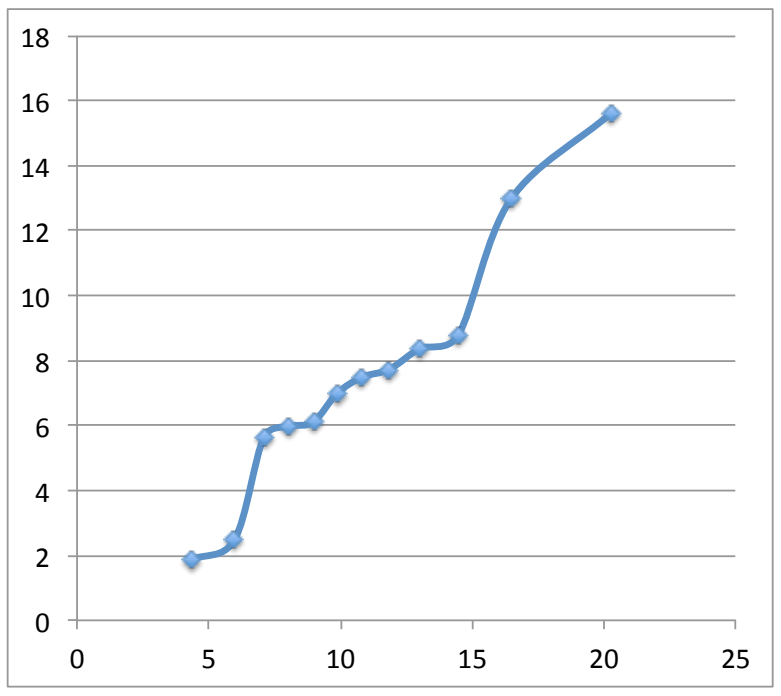

Figure 13. (a) A heat map of Cook's distances followed by a column with their sum per taxon. (b) A *Q-Q* plot of the sum of squared standardized residuals per taxon versus a chi-square of 11 degrees of freedom.

| | | | | | | | | | | | | | |
|---|---|---|---|---|---|---|---|---|---|---|---|---|---|
| ER3733 | 0.00 | 0.00 | 0.02 | 0.13 | 0.06 | 2.51 | 0.08 | 0.08 | 0.58 | 3.65 | 0.29 | 0.55 | 7.96 |
| ER3883 | 0.00 | 0.00 | 0.02 | 0.88 | 0.29 | 1.50 | 2.31 | 0.09 | 0.06 | 0.00 | 7.17 | 0.05 | 12.39 |
| Kabwe | 0.02 | 0.02 | 0.00 | 0.33 | 0.74 | 1.45 | 1.26 | 0.02 | 1.01 | 1.72 | 3.80 | 0.02 | 10.38 |
| Dali | 0.13 | 0.88 | 0.33 | 0.00 | 0.48 | 4.53 | 1.31 | 1.57 | 0.03 | 0.43 | 0.13 | 0.14 | 9.96 |
| LH18 | 0.06 | 0.29 | 0.74 | 0.48 | 0.00 | 0.88 | 2.26 | 0.19 | 0.21 | 0.12 | 2.75 | 0.66 | 8.65 |
| CHAPELLE | 2.51 | 1.50 | 1.45 | 4.53 | 0.88 | 0.00 | 0.00 | 3.81 | 0.20 | 1.55 | 1.43 | 0.02 | 17.89 |
| QUINA5 | 0.08 | 2.31 | 1.26 | 1.31 | 2.26 | 0.00 | 0.00 | 0.00 | 0.17 | 0.14 | 0.09 | 0.01 | 7.64 |
| IwoEleru | 0.08 | 0.09 | 0.02 | 1.57 | 0.19 | 3.81 | 0.00 | 0.00 | 0.60 | 1.18 | 3.05 | 0.07 | 10.67 |
| Khoi1F043H | 0.58 | 0.06 | 1.01 | 0.03 | 0.21 | 0.20 | 0.17 | 0.60 | 0.00 | 0.18 | 0.33 | 0.00 | 3.37 |
| Khoi1M044H | 3.65 | 0.00 | 1.72 | 0.43 | 0.12 | 1.55 | 0.14 | 1.18 | 0.18 | 0.00 | 2.71 | 0.18 | 11.86 |
| CROMAG1 | 0.29 | 7.17 | 3.80 | 0.13 | 2.75 | 1.43 | 0.09 | 3.05 | 0.33 | 2.71 | 0.00 | 1.09 | **22.83** |
| CROMAG3 | 0.55 | 0.05 | 0.02 | 0.14 | 0.66 | 0.02 | 0.01 | 0.07 | 0.00 | 0.18 | 1.09 | 0.00 | 2.80 |
| CROMAG1 | 0.00 | 1.43 | 7.17 | 2.71 | 3.05 | 3.80 | 0.13 | 2.75 | 0.29 | 0.09 | 0.33 | 1.09 | **22.83** |
| CHAPELLE | 1.43 | 0.00 | 1.50 | 1.55 | 3.81 | 1.45 | 4.53 | 0.88 | 2.51 | 0.00 | 0.20 | 0.02 | 17.89 |
| ER3883 | 7.17 | 1.50 | 0.00 | 0.00 | 0.09 | 0.02 | 0.88 | 0.29 | 0.00 | 2.31 | 0.06 | 0.05 | 12.39 |
| Khoi1M044H | 2.71 | 1.55 | 0.00 | 0.00 | 1.18 | 1.72 | 0.43 | 0.12 | 3.65 | 0.14 | 0.18 | 0.18 | 11.86 |
| IwoEleru | 3.05 | 3.81 | 0.09 | 1.18 | 0.00 | 0.02 | 1.57 | 0.19 | 0.08 | 0.00 | 0.60 | 0.07 | 10.67 |
| Kabwe | 3.80 | 1.45 | 0.02 | 1.72 | 0.02 | 0.00 | 0.33 | 0.74 | 0.02 | 1.26 | 1.01 | 0.02 | 10.38 |
| Dali | 0.13 | 4.53 | 0.88 | 0.43 | 1.57 | 0.33 | 0.00 | 0.48 | 0.13 | 1.31 | 0.03 | 0.14 | 9.96 |
| LH18 | 2.75 | 0.88 | 0.29 | 0.12 | 0.19 | 0.74 | 0.48 | 0.00 | 0.06 | 2.26 | 0.21 | 0.66 | 8.65 |
| ER3733 | 0.29 | 2.51 | 0.00 | 3.65 | 0.08 | 0.02 | 0.13 | 0.06 | 0.00 | 0.08 | 0.58 | 0.55 | 7.96 |
| QUINA5 | 0.09 | 0.00 | 2.31 | 0.14 | 0.00 | 1.26 | 1.31 | 2.26 | 0.08 | 0.00 | 0.17 | 0.01 | 7.64 |
| Khoi1F043H | 0.33 | 0.20 | 0.06 | 0.18 | 0.60 | 1.01 | 0.03 | 0.21 | 0.58 | 0.17 | 0.00 | 0.00 | 3.37 |
| CROMAG3 | 1.09 | 0.02 | 0.05 | 0.18 | 0.07 | 0.02 | 0.14 | 0.66 | 0.55 | 0.01 | 0.00 | 0.00 | 2.80 |

Figure 14. (a) A heat map of the squared externally studentized residuals followed by a column with their sum per taxon. (b) The previous heat map resorted by decreasing row sum.

While the sum of squares per taxon is interesting, it is an imperfect predictor of the fit of the model after a taxon is removed, partly due to tree search being invoked again. Indeed, looking at the sorted table of squared externally studentized residuals in figure 14b, it is noticeable that while CroMag1 and Chapelle fit the worst overall, they are neither highly incompatible with each other nor particularly similar in how they fit with other taxa. Thus it seems possible, that removing either one will not bring the other into good agreement with model expectations.

Jackknifing the taxa and repeating the tree search/parameter reveals the order of misfit (with the g%SD of the best tree reported followed by ig%SD in brackets) as CROMAG1: 3.672 (17.5617), CHAPELLE: 3.810 (20.7663), ER3883: 4.302 (22.8392), Kh1M044H: 4.238 (20.7933), Kabwe: 4.251 (20.5461), IwoEleru: 4.342 (20.2162), Dali: 4.190 (20.2601), LH18: 4.320 (20.8869), ER3733: 4.481 (22.79), QUINA5: 4.377 (23.5045), Kh1F043H: 4.660 (23.0775), and CROMAG3: 4.668 (22.7729). Here the order of fit is that predicted by the sum of squared residuals on the best tree. It is close to, but not identical to that in figure 14b. The order with the ig%SD fit is slightly different again. There appears to be a small jump in the magnitude of the misfit suggesting that CroMag1 and Chapelle are somewhat worse fitting than the others by g%SD, but by ig%SD only the former appears an outlier.

With this taxon Jackknifing, the tree is fully stable, except that when CroMag1 is removed, the two KhoiSan become sisters. It seems the large negative residual between the probable male pair of CroMag1 and the KhoiSan male is stabilizing the tree in an arrangement with the KhoiSan male sister to the pairing of KhoiSan female and the CroMag3 skull. In contrast, with CroMag1 removed and then Jackknifing again, the tree is again fully stable, except when Chapelle is removed that the KhoiSan female and CroMag3 again form a cherry. This interaction seems to be mediated by the positive residual of Chapelle and the KhoiSan male.

It is possible to interrogate the fit of data to model further, for example by using partial leverage and partial residuals. The former allow the contribution of each independent variable (in this case edge lengths) to the leverage of an observation to be assessed. However, the returns are often decreasing and they do not take into account model selection via tree search, which can have a strong effect.

### 3.11 Likelihood-based tests on the tree

It is also possible to make various tests and tree evaluations with likelihood. As mentioned earlier, since the g%SD is a rescaled likelihood, then going back to the likelihood or the log likelihood is straightforward. An interesting point arises when working with the likelihood itself. It is usually not important when evaluating a likelihood difference, but it is relevant to working with the number of fitted parameters. The likelihood can be evaluated with a predefined variance, or the empirical variance from the residual sum of squares can be used. The latter is an extra parameter estimate that reduces the degrees of freedom by 1 compared to if it is pre-specified. Thus, when basing certain statistics on the estimate likelihood itself, such as AICc, which care about how many parameters have been optimized, this is important. Thus, the g%SD itself is also reported with this extra degree of freedom lost when the variance is estimated directly from the residuals.

It is possible to compare an *a priori* hypotheses in the form of trees by comparing log likelihoods. Assuming a Chi-square distribution under the null hypothesis is usually safe as long as the hypotheses are nested, boundaries are not a major issue and the trees are well defined *a priori*.(e.g., Swofford et al. 1996, Ota et al. 2000). For the set of 12 taxa left after Qafzeh6 is removed, many parts of the tree are precisely as would be *a priori* predicted, such as the two erectus skulls together, or the two Neanderthals together. One *a priori* assumption interesting to test is the belief that Kabwe and Dali both represent a middle Pleistocene species that goes by a range of names such as *H. rhodesiensis* or *H. heidelbergensis*. If we accept that species should not be paraphyletic or polyphyletic, then an appropriate test here is to compare the likelihood associated with the best tree with the constraint that Kabwe and Dali are sister taxa to that with no constraint. This involves two maximizations of the likelihood over different sets of trees, the effect of which should really be addressed by a simulation. In this case a fudged test can be made since, on the optimal tree, they are only one edge apart (something that could not be known *a priori*). Thus, the constraint of monophyly sees that edge collapsed to zero and a common edge free to take on a positive length (in this case it sits on the boundary with length zero). Ignoring the maximization effects, we know the difference of likelihood to expect under the null model and one that has a single internal edge collapsed. This test on a boundary is described by Ota et al. (2000).

Doing the test in this fashion described above has the g%SD under $H_0$ (monophyly) being 4.415% ($k$ = 21) and that under $H_1$ being g%SD = 4.335 ($k$ = 22). Since $g\%SD = \left(\frac{1}{N-k}\right)^{0.5} \exp\left[\frac{-\ln L}{N}\right] \times C$, the difference in log likelihood, equals the log of the likelihood ratio, can be calculated as $N(\{\ln[(zg\%\mathrm{SD}(\mathrm{H}_0)]\text{-}\ln[g\%\mathrm{SD}(\mathrm{H}_1)]\}) = N(\{\ln[(zg\%\mathrm{SD}(\mathrm{H}_0)/g\%\mathrm{SD}(\mathrm{H}_1)]\})$, where $z = \sqrt{(N-k_{H0})/(N-k_{H1})}$ is the adjustment if $k$ is different. In this case the $\Delta \ln L$ = 1.948. Twice this value or 3.897 is distributed as a 0.5:0.5 mixture of $\chi_0^2$ (a delta function at zero) and $\chi_1^2$ under the assumptions made here (Ota et al. 2000). This is the same as asking what is half the probability of a value of 3.897 or larger from a $\chi_1^2$ distribution, which is 0.0484/2 = 0.0242. Thus if the test was set up at a p value of 0.05 the null hypothesis would be rejected, that is, the monophyly of Dali and Kabwe is rejected.

Another way of looking at this test which favors $H_0$ and better takes into account the local maximization by tree search when that part of the tree is not specified *a priori*, is to consider the range of alternatives to $H_0$. If $H_0$ does indeed sit on the boundary, then there are three resolutions of this four way internal node, each following the mixed chi-square marginal distribution mentioned above. One of these is an edge in favor of monophyly, which is disregarded since we are already on the boundary for this data/model and the other two include the edge in the optimal tree and the other would reverse the order of Kabwe and Dali. Assuming that the correlation of these two edge lengths is zero, then, in that case, the situation is like that described in Waddell et al. (2002), inducing a 25% zero plus 75% exp(-x) distribution for the log likelihood difference. Making the test this way, the observed p value under $H_0$ is 0.107 and the null hypothesis is not rejected.

Another *a priori* null hypothesis might be that Iwo Eleru is a modern human skull. On the 12 taxon example this effectively requires Iwo Eleru to locate somewhere within the clade of the four modern humans. This test is reasonable, once it is recognized that CroMag1 tends to be about as deeply divergent from other modern skulls as should be expected, even in larger sample. Since the trees being compared are not nested, it is safer to go with the distance analogy to the Kishino and Hasegawa (1989) (KH) tests or the AU test (Shimodaira and Hasegawa 1989). The KH test compares two trees based on the log likelihood scores of each of $c$ site patterns.

For the KH test applied to distance data, the log likelihood difference of a distance fitted to the tree replaces a site pattern log likelihood. While there are only $t(t\text{-}1)/2$ informative distances, as long as this is larger than ~ 50, it should be reasonable. Further, use of a winning distance/sign test instead of a paired log likelihood z or t test, should be robust to giving false positives with small sample sizes. Comparing two trees with the KH test is also analogous to a bootstrap without tree search where just two trees are compared.

The result of testing the best tree with Iwo Eleru within the clade of modern taxa with the ML tree is absolutely rejects IwoEleru as a modern human. The log likelihood of each distance fitted onto the tree is, in this case, ignoring constants, $\ln L(D_{obs_i}) = -0.5\ln(\sigma^2) - 0.5(D_{obs_i} - D_{\exp_i})^2 / \sigma^2$, where $\sigma^2$ is for that tree being considered.

The g%SD of the ML tree, as before is 4.335 ($k$ = 22), while that on the best Iwo Eleru modern tree is 10.472 ($k$ = 21). The best Iwo Eleru modern tree has a polytomy branching pattern (CroMag1,IwoEleru,(KhoiSanMale,(KhoiSanFemale,CroMag3))). The log likelihood ratio difference is 58.953. Because the variance is so much higher on the IwoEleru with modern's tree, every single observed distance on the ML tree has a markedly higher ln$L$. Thus the winning distances statistic is 63 wins with 3 draws for the shared tree cherries. Null hypothesis of equal numbers of winning distances are completely rejected. The paired $t$-test similarly gives a 2-tailed probability p of 1.8e-109 for the two sets of site log likelihoods being equal.

This test does, however, has some unfairness towards $H_0$ of IwoEleru being a modern skull. This is because the ML tree still has a bit more freedom in terms of rearrangements to minimize the total sum of squares. This concerns, in particular, the three possible arrangements of IwoEleru and LH18 relative to the other skulls. One of these, which is not optimal, is IwoEleru sister to all the moderns. One way to combat this bias against $H_0$ is to force IwoEleru to fall in this position when doing the ML search for $H_1$. Doing a search with this constraint in place yields the same tree as the ML tree except that the short internal edge between LH18 and IwoEleru is collapsed to zero. The new g%SD is 4.550 ($k$ = 21). The ln$L$ difference is reduced to 55.02, but the winning sites test is the same and the paired two tail $t$-test barely changes with p = 6.9e-109. Thus, $H_0$ is still more than convincingly rejected.

It is also possible to redo these tests allowing negative edge lengths. In general this will tend to favor the worse fitting hypothesis, by bringing into consideration models that are ruled out *a priori*, and thus simply dropping the power of the test for no obvious gain.

While the likelihood ratio statistic is not nested, assuming $H_0$ is a polytomy of the form (CroMag1,IwoEleru,(KhoiSanMale,(KhoiSanFemale,CroMag3))), and the orthogonality of edge lengths, then the difference in log likelihood about one expansion of an internal node should be 25% zero and 75% following an exponential distribution (Waddell et al. 2002). More particularly, such a pair of edges, produced under the null hypothesis, of no resolution of Iwo Eleru from modern humans, should produce a ln$L$ difference of plus minus 2.705 ln$L$ 95% of the time. The Iwo Eleru modern and the Iwo Eleru sister to all moderns induce just such a pair of edges. On this data, ln$L$ difference is so far outside this confidence interval (55.02 versus +/-2.705 expected) this hypothesis is rejected. Using the stated distribution of such a pair under this null hypothesis from Waddell et al. (2002), the probability of finding an observation from an exponential of mean 1 that is at 55.02 is just 1.27e-24. Since ¼ of the actual distribution is at zero, then this should be adjusted to 0.75 times 1.27e-24 or 9.55e-25. Thus the null hypothesis that Iwo Eleru could have come from inside the clade of moderns in this sample of data is again rejected with extreme prejudice.

Finally, it is interesting to see how these tree models of skull shape treat the multiregional hypothesis (Wolpoff 1999), that is, that there was genetic continuity in each region of the world since the time of *Homo erectus*, thus modern Europeans evolved from European *Homo erectus* and modern Africans evolved from African *Homo erectus* (e.g., Klein 2009). Against this is the optimal unconstrained tree, which shows a pattern of multiple lineages emerging and apparently replacing others over the last 2 million years, while the closest relatives of modern humans occur in Africa over the period since Neanderthals diverged, roughly the last 400,000 years. The data set here is trimmed slightly to make the test clear. Dali is removed as there are no later Asians in this sample, while Qafzeh6 is removed on account of its intermediate geographic locality, and also being unclear if it left genes in modern populations. For the unconstrained tree ($T_1$) the values were, g%SD ($P$ = 0, $k$ = 1) = 3.421 and optimal with $P$ and $P$' free was g%SD ($P$ = 2.240, $k$ = 1) = 3.335. The corresponding numbers for a tree with the African skulls and the European skulls to be monophyletic ($T_2$) were g%SD ($P$ = 0, $k$ = 1) = 18.063 and optimal with $P$ and $P$' free was g%SD ($P$ = -7.755, $k$ = 1) = 8.383. In order to evaluate an Out of Africa model with minor genetic leakage, a constraint of the reverse of the multiregional hypothesis was imposed (archaic African skulls with the European Cro-Magnons and archaic European skulls with modern

Africans, $T_3$). The results here were g%SD ($P$ = 0, $k$ = 1) = 18.063 and optimal with $P$ and $P$' free to vary was g%SD ($P$ = -7.2.66, $k$ = 1) = 9.368.

It is not possible to make a simple likelihood ratio test in this situation, since the resulting trees are not nested. Another possibility would be a KH-like test of the trees likelihood values as done above. A third possibility is to explore the posterior probability of each model using the BIC approximation to the posterior probability of the model used in Waddell et al. (2002). This assumes equal priors for the three tree models, which seems reasonable, as all have potentially the same number and types of parameters. Since g%SD for the same data is inversely monotonic with the likelihood of the data per distance, it is tempting to make the comparison as 1/g%SD($T_1$)/{(1/g%SD($T_1$) + 1/g%SD($T_2$)} : 1/g%SD($T_2$)/{(1/g%SD($T_1$) + 1/g%SD($T_2$)}, yielding a posterior ratio of 0.725: 0.137. However, this would be analogous to the differences in political and scientific views of the Star Wars project (rumored as a groups of scientists reporting to President Reagan that they needed a laser with energy of the order of 10 to the power 40, but after years of research they had only reached an order of 10 to the power of 22, to which the President encouraged them to have confidence and report back to him next year as they were over half way there!). The g%SD values are proportional to the inverse of $\exp(\ln L / N)$, so the comparison should be 1/g%SD$(T_1)^N$/{(1/g%SD$(T_1)^N$ + 1/g%SD$(T_2)^N$}, etc., which at $P$ = 0, yields a ratios of 1 versus 1.2e-40. With $P$ and $P$' free to take their optimal values the ratios are 1 to 9.6e-23. The multiregional model looks very very unlikely.

Thus the multiregional hypothesis does very poorly compared to the Out of Africa total replacement model. The posterior probabilities of $T_2$ and $T_3$ are identical at $P$ = 0, but with $P$ and $P$' free, their ratios to each other become ~ 0.998 to 0.002 or a relative posterior probability of about 450 to 1 in favor of some genetic leakage. Thus, a genetic leakage Out of Africa model may find some support relative to a total replacement model, although the $P$ and $P$' values to do this were extreme (e.g. $P$ = -7.7). The relative support for a genetic leakage model should be examined in future on a larger set of skulls.

### 3.12 Likelihood and fit statistics with distance transformations

In the materials and methods section it was mentioned that the square root of the Procrustes distance was used, rather than the sum of squared deviations, which is the Procrustes distance. The reason for this is that various fits are far better when using the square root rather than using the Procrustes distance itself. To show how this is done, it is necessary fist to calculate the ln$L$ of the data with transformation.

Usually, within a likelihood framework, the expected variance-covariance matrix of the data after a transformation can be assessed, using a delta-method approximation. This is taking the first derivative at each point the data is transformed at (Stuart and Ord 1987, 1990). Here, let $q$ be the power the distances are raised to, that is $\mathbf{D}_{obs} = (\mathbf{D}_{Proc})^{q}$, where "Proc" stands for Procrustes. Thus, $q$ potentially becomes another free parameter. With ordinary least squares (OLS), where the variance is estimated from the sum of squared residuals, the likelihood of the data becomes $\prod_{i=1}^{N} \frac{1}{\sqrt{2\pi\sigma^2}} \exp\left[\frac{-\left(D_{obs_i} - D_{\exp_i}\right)^2}{2\sigma^2}\right] q\left(D_{\mathrm{Proc}_i}\right)^{q-1}$, where the last term adjusts for the transformation by power $q$ by taking the first derivative (gradient) at each data point where the transformation is applied. Cross derivatives are not needed as the data points are assumed to be independent. This work falls in the framework of a general linear model, where $g(\mathbf{D}_{\mathrm{obs}}) = (\mathbf{D}_{\mathrm{obs}})^q$ is considered the link function.

To illustrate how this transformation works, recall that the variance of a variable multiplied by $x$ increases $x^2$. Thus in transforming variances the gradient at a data point squared is used, while to correct the covariance of $i$ and $j$, the product of the gradient at $i$ and at $j$ is used, for example, in transforming phylogenetic site pattern counts into "tree edge length space" (Waddell et al. 1994). Consider the case of $x = 2$, so the variance becomes 4 times its previous value. Thus, we have $Uncorrected_Likelihood(2\mathbf{D}_{obs}) = \prod_{i=1}^{N} \frac{1}{\sqrt{2\pi 4\sigma_i^2}} \exp\left[\frac{-\left(2D_{obs_i} - 2D_{\exp_i}\right)^2}{2\times 4\sigma_i^2}\right] = \prod_{i=1}^{N} \frac{1}{2}\frac{1}{\sqrt{2\pi.\sigma_i^2}} \exp\left[\frac{-4\left(D_{obs_i} - D_{\exp_i}\right)^2}{2\times 4.\sigma_i^2}\right]$.

The multiplication effect inside the exp function cancels, while the overall effect per data point was to halve the likelihood. Thus multiplying each data point's likelihood by the first derivative, of 2 corrects the likelihood to what it should be.

For unweighted weighted least squares, with independent observations, and assuming $P$ and $q$ are independent, the likelihood of the data after the transformation of $D_{obs_i} = (D_{Proc_i})^q$ can be written $\prod_{i=1}^{N} \frac{1}{\sqrt{2\pi\sigma_i^2}} \exp\left[\frac{-\left(D_{obs_i} - D_{\exp_i}\right)^2}{2\sigma_i^2}\right] q\left(D_{\mathrm{Proc}_i}\right)^{q-1}$, where the variance of each data point is used. The variance at each observation can be of the form $\sigma_i^2 = w_i\sigma^2$, where $\sigma^2$ can be a global or pooled estimate of the variance. Here, the ML estimator of $\sigma^2$ is $\frac{1}{N}\sum_{i=1}^{N}\frac{\left(D_{obs_i} - D_{\exp_i}\right)^2}{w_i}$, the weighted sum of squares divided by $N$, denoted WSS/$N$. In practice, a general form of the weights, for example $w_i = \left(D_{obs_i}\right)^P$, which are power weights, or $w_i = \exp\left(P'\times D_{obs_i}\right)$, which are exponential weights, are often used. The likelihood of the data thus becomes $\prod_{i=1}^{N} \frac{1}{\sqrt{2\pi w_i\sigma^2}} \exp\left[\frac{-\left(D_{obs_i} - D_{\exp_i}\right)^2}{2w_i\sigma^2}\right] q\left(D_{\mathrm{Proc}_i}\right)^{q-1}$

Taking the natural logarithm of this equation, the ln$L$ of the data is

$$\sum_{i=1}^{N}\left(-\frac{1}{2}\ln[2\pi] - \frac{1}{2}\ln[w_i] - \frac{1}{2}\ln[\sigma^2] - \frac{\left(D_{obs_i} - D_{\exp_i}\right)^2}{2w_i\sigma^2} + \ln[q] + (q-1)\ln[D_{\mathrm{Proc}_i}]\right).$$

$$= -\frac{N}{2}\ln[2\pi] - \frac{1}{2}\sum_{i=1}^{N}\ln[w_i] - \frac{N}{2}\ln[\sigma^2] - \sum_{i=1}^{N}\frac{\left(D_{obs_i} - D_{\exp_i}\right)^2}{2w_i\sigma^2} + N\ln[q] + (q-1)\sum_{i=1}^{N}\ln[D_{\mathrm{Proc}_i}]$$

Next the WSS/$N$ is substituted in, first in one place then in another,

$$= -\frac{N}{2}\ln[2\pi] - \frac{1}{2}\sum_{i=1}^{N}\ln[w_i] - \frac{N}{2}\ln[WSS/N] - \frac{1}{2\sigma^2}\sum_{i=1}^{N}\frac{\left(D_{obs_i} - D_{\exp_i}\right)^2}{w_i} + N\ln[q] + (q-1)\sum_{i=1}^{N}\ln[D_{\mathrm{Proc}_i}]$$

$$= -\frac{N}{2}\ln[2\pi] - \frac{1}{2}\sum_{i=1}^{N}\ln[w_i] - \frac{N}{2}\ln\left[\frac{1}{N}\sum_{i=1}^{N}\frac{\left(D_{obs_i} - D_{\exp_i}\right)^2}{w_i}\right] - \frac{1}{2WSS/N}WSS + N\ln[q] + (q-1)\sum_{i=1}^{N}\ln[D_{\mathrm{Proc}_i}]$$

$$= -\frac{N}{2}\ln[2\pi] - \frac{1}{2}\sum_{i=1}^{N}\ln[w_i] - \frac{N}{2}\ln\left[\frac{1}{N}\sum_{i=1}^{N}\frac{\left(D_{obs_i} - D_{\exp_i}\right)^2}{w_i}\right] - \frac{N}{2} + N\ln[q] + (q-1)\sum_{i=1}^{N}\ln[D_{\mathrm{Proc}_i}]$$

and following cancelling of the WSS terms, the final form is arrived at. Also, noting that the last summation can be expressed as $N$ times the natural log of the geometric mean ($g$) of the informative Procrustes distances, leads to the last two terms being expressed as $N\left\{\ln[q] + (q-1)g(\mathbf{D}_{\mathrm{Procrustes}})\right\}$.

An important point to make here, that can otherwise case cause considerable confusion, is that the weights are applied to the variance. That is, they are part of the error structure of the model. They then act inversely upon the fitting of individual data points, since this is the way to obtain the maximum likelihood estimates of parameters under that model's assumption. Sometimes, in phylogenetics, the weights are seen as multipliers on fitting of the individual data points, without acknowledging that this comes about as part of the model of the error structure. In this case, they are in fact the inverses of the weights being used here. Herein, the weights are used consistently in terms of model-based predictors of the scalar on errors of data points.

The usual g%SD statistic is derived by first taking the log likelihood for weighted least squares without considering transformations, dividing through by $N$, exponentiating, and changing sign, to give $\exp\left[\frac{-\ln L}{N}\right] = C \times \sqrt{geomean(w_i)} \times \sqrt{\frac{1}{N}\sum_{i=1}^{N}\frac{\left(D_{obs_i} - D_{\exp_i}\right)^2}{w_i}}$. This quantity is inversely monotonic with the log likelihood. Ignoring the term $C$, which is constant with respect to the number of taxa and hence informative distances (or the number of data points), plus adjusting the sum of squares to a less biased form by replacing $1/N$ by $1/(N\text{-}k)$, then dividing by the geometric mean of the observed distances and finally, multiplying by 100% gives $g\%SD = \frac{\sqrt{geomean(w_i)}}{geomean(\mathbf{D}_{obs})} \times \sqrt{\frac{1}{N-k}\sum_{i=1}^{N}\frac{\left(D_{obs_i} - D_{\exp_i}\right)^2}{w_i}} \times 100\%$, as a measure of the additivity of distances on a tree. The g%SD measure acts like a coefficient of variation (CV) measure of fit.

The $\ln L$ of the data, without the constant terms, is recovered as $\ln L = -N \ln\left[\frac{g\%SD}{100\%} \times geomean(\mathbf{D}_{obs}) \times \sqrt{\frac{N-k}{N}}\right]$. The full $\ln L$ is therefore this last value plus the constant term $-\frac{N}{2}\left(1 + \ln\left[2\pi\right]\right)$.

When using a measure of fit such as g%SD there is use of the inverse of the mean of the distances to adjust the form of the geometric mean of the likelihood per data point into an error form that seems intuitive, and is analogous to the percentage of error in a single measurement. However, with certain transformations, such as $q$ goes to zero, then all distances become equal (in this case 1). This in turn means that a star tree with all terminal edge lengths equal will fit the data perfectly. However, this “dark star” of a tree is also effectively eroding the useful information in measures such as g%SD. Another correction that may be made is to divided the g%SD by the sum of the internal edge lengths of the tree being considered, divided by the total sum of edge lengths, to yield the ig%SD measure. This is potentially useful as the resolution of the internal edges is typically the aspect of the tree researchers are most interested in and it also means the ig%SD becomes increasingly poor as a star tree is approached.

The g%SD measures are inversely monotonic with average log likelihood per data point as long as the data are not transformed. If a known transformation is applied to the data, the new likelihood can be estimated as described above. A real benefit of using g%SD like measures is in gaining a better intuitive appreciation of how well distances fit a tree, which implies some degree of intuitive comparison between different data sets and different models, something likelihood is less suited to do if an unknown scale factor or transformation is involved. In this regard, they can act somewhat like the consistency and retention indices for character data fitted to a parsimony phylogenetic tree.

A further set of fit measures worth considering, use the proportion of the variance explained, and are known as coefficients of determination. For example, $R^2(OLS) = 1 - \frac{SS_{res}}{SS_{tot}}$ or its adjusted, less biased, form $R^2(OLS) = 1 - \frac{SS_{res}}{SS_{tot}}\frac{N-m}{N-k}$, here where $N$ is the number of informative distances, $k$ are the fitted parameters used in estimating the residual sum of squares ($SS_{\text{res}}$) and $m$ are the fitted parameters used in estimating the total sum of squares ($SS_{\text{tot}}$). Consider two ways to estimate the total sum of squares. The first calculates the sum of squares total as the variance about what, in simple linear regression, is a flat line with just an intercept term equal to the mean of the data points, and no slope. That is, $SS_{tot} = \sum_{i=1}^{N}\left(D_{obs_i} - \bar{D}_{obs}\right)^2$, where the second term is the arithmetic mean of the informative distances. The second approach uses $SS_{tot} = \sum_{i=1}^{N} D_{obs_i}^2$. The dark star mentioned earlier is composed of a star tree with all edge lengths equal to the mean of the informative distances divided by two, so by incorporating deviations from the mean, the dark star might be nullified as a strong attractor.

With weighted least squares, one analogue of $R^2$ is $R^2(WLS) = 1 - \frac{WSS_{res}}{WSS_{tot}}$, the weighted sum of squares is the same as that optimized by the likelihood criterion, as used above and previously. For a given value of *q*, *P* is optimized by PAUP when reporting $WSS_{res}$. The term $WSS_{tot}$ may be defined as $WSS_{tot} = \sum_{i=1}^{N} \frac{\left(D_{obs_i} - \bar{D}_{obs}\right)^2}{w_i}$, where $\bar{D}_{obs} = \frac{\sum_{i=1}^{N} w_i D_{obs_i}}{\sum_{i=1}^{N} w_i}$. However, if of the form, $w_i = D_{obs_i}^P$, it is not clear which value of *P* should be used. If the value of *P* for $WSS_{res}$ is used, then the value of $WSS_{tot}$ has not been minimized and how far it is from the true value of $WSS_{tot}$ might also fluctuate across *q*. The optimization of *P* for $WSS_{tot}$, in terms of likelihood, can use the ln*L* formula for *WSS* given above.

Finally, there is a appealing generalization of $R^2$ (Magee 1990) which is defined for a tree as

$$R^{2(L)} = 1 - (L_{DS} / L_T)^{2/N} = 1 - \left(\exp\left\{\ln L_{DS} - \ln L_T\right\}\right)^{2/N} = 1 - \left(\exp\left\{\left(\ln L_{DS} - \ln L_T\right)/N\right\}\right)^2$$

$$= 1 - \left(\exp\left\{\ln L_{DS}/N - \ln L_T/N\right\}\right)^2 = 1 - \left(\exp\left\{-\ln L_T/N\right\}/\exp\left\{-\ln L_{DS}/N\right\}\right)^2 = 1 - (g\%SD_T / g\%SD_{DS})^2$$

where $L_{DS}$ is the likelihood of the dark star tree, $L_T$ is the likelihood of the other tree under consideration, and the g%SD measures are made assuming the same number of estimated parameters for both trees. Thus, maximizing this statistic on a given tree is the same as maximizing the likelihood ratio of the tree under consideration versus a dark star tree (it being a tree that can be otherwise be strongly favored as a byproduct of data transformation). If each g%SD is made using an adjustment of 1/(*N*-*k*), where *k* is specific to that tree, then the result is a generalization of $R^{2(L)}$ to the adjusted for the degrees of freedom form of $R^2$. This also makes it clear that the g%SD can be adjusted by using the root mean square of the observed distances less their mean, in place of the geometric mean of the distances, to have a g%SD value that will not go to zero on the dark star tree.

One last statistic that is favored for assessing general linear models is the Pearson correlation statistic for the observed and the expected data (Zheng and Agresti 2000). For distances, this should be made excluding the zeros on the diagonal of the data matrix, since they are not considered part of the data (their inclusion increases the correlation towards one as an increasingly large constant is added to all the informative distances).

### 3.13 Likelihood and fit after transforming the distance

The fit by different methods as $q$ ranges from 0 to 1.5 is evaluated and shown in figure 15. The log likelihood of the transformed data with $P$ set to zero is calculated as described above and there is clearly a peak very close to $q$ = 0.5 as seen in figure 15a. To assess if the log likelihood at the optimal $q$ value is significantly higher than that at $q$ = 1, a two-tailed likelihood ratio test with one degree of freedom is appropriate since, *a priori*, we have not specified if the optimal $q$ should be larger or smaller than 1. First though, the difference in likelihood between $q$ = 0.52 and 0.5 is ~0.05 ln$L$ units this is not significant, and the square root was used as it is a commonly used transformation. The difference in log likelihood from $q$ = 1 is 16.13 ln$L$ units which is highly significant (p << 0.0001). Even allowing for $P$ to be freely optimized, the difference in ln$L$ between the peak near $q$ = 0.56 and $q$ = 1 is 3.96, with an associated p of 0.005, so again the null hypothesis of $q$ = 1 is decidedly rejected. The shape of the ln$L$ curve with $P$ free and $q$ > 0.8 suggests that $P$ and $q$ might be significantly interacting parameters.

Figure 15b shows how various g%SD measures fit the data as $q$ varies. As already noted, g%SD measures are not intended to replace properly transformed ln$L$ measures when available. As can be seen, g%SD measures are strongly attracted to the dark star tree and show only one minimum, that is, when it is reached. The g%SD measures for a given $q$ do still rank the likelihoods of the models, and it can be seen that optimizing $P$ rather than $P$' tends to produce a better fit most of the time with this data. The orange and red line of figure 15b track the optimal value of $P$ and $P$' respectively. These cross at $P$ = 0 at a value of $q$ slightly lower than 0.5. Both are negatively correlated with $q$, and the relationship in the range $q$ ~ 0.5 - 1 seems roughly linear.

The ig%SD measures with $P$ = 0 and $P$ optimized both have minima at q ~ 0.7 (figure 15b). While this is not as low of a $q$ value as with likelihood, it is still clearly lower than at $q$ = 1. These are intended as flexible measures of additivity and the resolution of internal edges in a distance tree, so it is encouraging that they do coincide moderately well with likelihood in example. If anything, they might seem to be pushing the optimum too far away from the dark star tree.

Finally, figure 15c shows the optimal fit suggested by various versions of the $R^2$ coefficient of determination measure. With $P$ = 0, both the raw and the adjusted measures have clear peaks at $q$ ~ 0.55. These measures have a term related to the sum of squares for a star tree with all edges of an optimal equal length, and none are attracted to the dark star tree in this example. The bump in the adjusted $R^2$ near $q$ = 1.1 is due to the length of an external edge in the tree (that leading to LH18) hitting the boundary at 0, that is, it must be non-negative, and at that point this reduces the penalty on the number of fitted parameters associated with the sum of squares residual. The $R^2$ values associated with the weighted sum of squares suggest a minimum at $q$ near 0.57. However, their fit by $R^2$ is not as smooth and there is the suggestion of other optima at $q$ greater than 1. In the case of the adjusted $R^2$ for weighted least squares fitting, the very highest point is near $q$ = 1.2. However, the weighted least squares statistic $R^2$ is being affected by two then three external edge length parameters going to zero a $q$ increases.

For this data, the optimal value with $R^{2(L)}$ falls at $q$ 0.63 with a value of 0.983 unadjusted. This is a very high value for this statistic and confirms that the optimal ML binary tree with $P$ and $q$ also fitted explains most of the variance in this data. It is useful to consider how many parameters, $k$, are estimated in g%$S_{DT}$ and g%$SD_{DS}$. For a full binary tree we have 2$t$-3 edge length parameters, plus parameter $P$, and the variance. For this data from $t$ = 12 taxa, that is 23. For the dark star tree with all edges of equal length, there is the estimation of the weighted mean of the data, which divided by 2 gives the edge lengths, and also of the variance for that model, so $k$ = 2.

For this data with $P$ and $q$ optimized, the optimal correlation of the observed and the expected distances was at $q$ ~ 0.53, in close agreement with the maximum of the basic

$R^2$ statistic.

Fully delving into model selection with $P$ and $q$ both free, plus trees of various degrees of resolution is beyond the scope of this article. However, it does require consideration of many submodels on distinct binary trees close to the optimal binary tree, which, on this example, did not change as either $P$ or $q$ were varied, except with $q$ 1.3 or greater. At that point three external and one internal edge (that linking the Neanderthals) collapsed to zero (with the Neanderthals then becoming paraphyletic to each other).

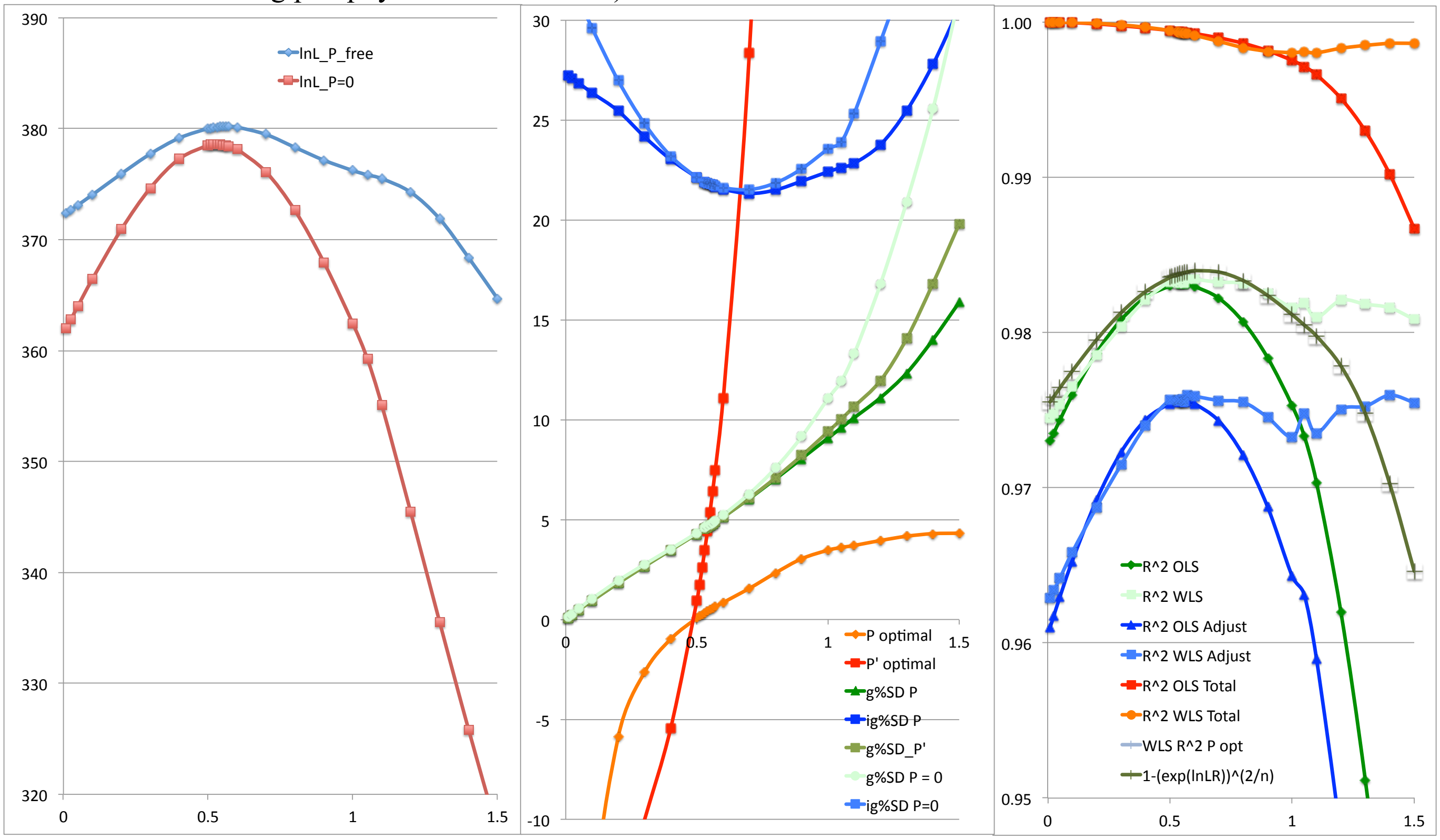

Figure 15. The optimal fit for parameter $q$ in the range of 0 to 1.5, with 1 being the standard Procrustes distance. (a) Leftmost, the fit according to the log likelihood for the data ($y$-axis) transformed by power $q$ ($x$-axis). The optimal value falls at $q \sim 0.52$ when $P$ is fixed at 0, and when $P$ is fitted to be optimal, the highest log likelihood occurs at $q \sim 0.56$. (b) Center, the fit with respect to a range of g%SD measures, along with the associated optimal values for $P$ and $P'$, all on the $y$-axis and power $q$ on the $x$-axis. The g%SD measures are shown in green, while the optimal values for $P$ and $P'$ are shown in orange and red. The implied optimum is for very negative $P$ and $P'$ values when $q$ goes to zero. The internal or ig%SD measures for $P$ fixed at zero or flexible, are shown in blue. Their optimal values fall at $q \sim 0.68$ and 0.71 respectively. (c) Rightmost, the fit using $R^2$ statistics for a range of $q$ values. The optimal values for $P = 0$ and $R^2$ OLS and $R^2$ OLS adjusted by residual degrees of freedom both fall at $q \sim 0.55$. The $R^2$ WLS and $R^2$ WLS measures have multiple optima, with the highest optimum at $q \sim 0.57$. The olive green line is the statistic $R^{2(\mathrm{L})}$ or the likelihood ratio generalization of $R^2$. The red and orange lines show $R^2$ using the total sum of squares without fitting the mean of the informative distances.

Counting the number of parameters in some of these models also requires further consideration. While $q = 1$ is the null hypothesis for the Procrustes distance, the null hypothesis for $P$ is less clear. In one sense, $P = 0$ is a naturally favored null hypothesis for the distribution of errors with respect to the data points, in that it implies the simpler homoscedastic assumption. On the other hand, if skull shape is neutral and like variables such as height, that is, influenced by the frequencies of many dozens or hundreds of genes, the Procrustes distance may be expected to behave approximately as a Brownian walk on a tree. In this case, its variance should be proportional to the distance between skulls, that is, $P = 2$. Repeating the earlier likelihood ratio hypothesis test, with the null hypothesis that $q = 1$, $P = 2$, versus $P$ and $q$ allowed to take on their optimal values, the ln$L$ difference if 6.43 and p = 0.0016. The rest of these two models are also fully nested (the optimal trees are identical), so we again reject the null hypothesis that $q = 1$ in favor of $q = 0.5$, with considerable confidence.

The correlation of $P$ and $q$ can be seen in the approximation that occurs when $e$, the error on $D_{obs}$ is proportional to $D^q_{obs}$. That is $D^q_{obs_i} = D^q_{\exp_i} + error = D^q_{\exp_i} + D^q_{\exp_i} e$, and $e$ is small compared to $D_{\exp}$ so that $D_{obs} \sim D_{\exp}$, to give $\frac{\left(D^q_{obs_i} - D^q_{\exp_i}\right)^2}{w_i \sigma^2} = \frac{\left(\left(D^q_{\exp_i} + D^q_{\exp_i} e\right) - D^q_{\exp_i}\right)^2}{D^P_{obs_i}\sigma^2} = \frac{\left(D^q_{\exp_i} e\right)^2}{D^P_{obs_i}\sigma^2} \approx \frac{D^{2q}_{obs_i} e^2}{D^P_{obs_i}\sigma^2}$.

Thus, adjusting $P$ can cancel some, but not all of, the effect, of altering $q$. Even if the true errors are independent of the size of $D_{obs}$, then the effect of trying to fit $(D_{obs})^q$ additively onto a tree can induce errors positively or negatively correlated with the size of $D_{obs}$, with the outcome that $P$ and $q$ often covary in a non-linear way, but in the same direction, that is, as $q$ gets bigger so to does $P$. Because the exponential weights model can yield weight functions similar in shape as $P$' varies to that of the power model, then $P$' and $q$ will also covary, as seen in this example. While fitting exponential weights to binary trees here was slightly worse than fitting power weights, this is not always the case and the reverse is often seen.

### 3.14 The Procrustes, $q = 1$, $P = 2$, the neutral genetic model

As just seen, taking the Procrustes distances to the power of 0.5 increases the likelihood of the model. However, it is still useful to look at the predictions of the Procrustes distance, that is $q = 1$. In particular, examination with $P = 2$ is desirable, as this is the power that is matched to a Brownian walk on the tree for this distance. This distance and variance also tie in exactly with the assumptions of a fully neutral model where shape variations are due to random fluctuations in many genes each exerting a small influence on the overall skull shape.

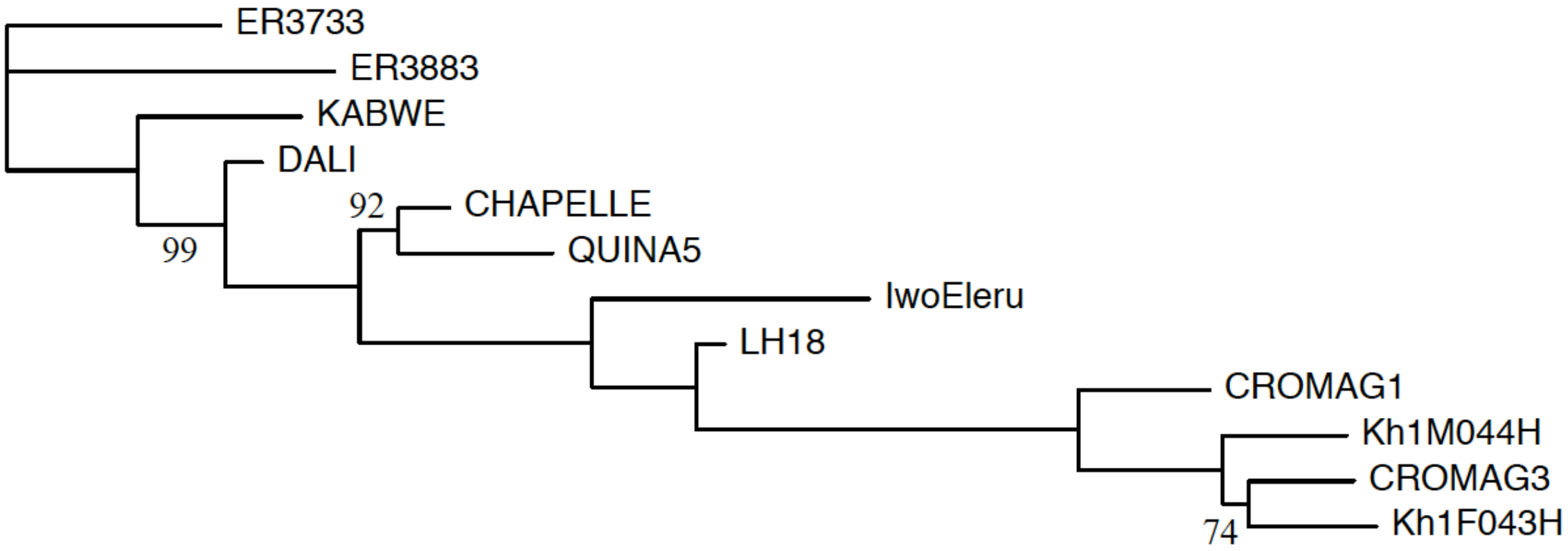

Figure 16. The optimal weighted tree recovered from Procrustes distances ($q = 1$) and $P = 2$ with a non-negativity constraint on edge lengths. If the residual resampling support of an internal edge from 10,000 replicates rounded was less than 100, it is shown.

The weighted tree for the Procrustes distance is shown in figure 16. Note that it strongly reinforces the earlier view, from the root Procrustes OLS+ tree, that the rate of evolution has been accelerating down the lineage leading to modern humans. This is emphasized by that fact that with respect to the scale of the overall tree (nearly 2 million years from root to modern tip), the tips marked by Chapelle, Quina5, Iwo Eleru, CroMag1 and 3, and Kh1 male and female, are effectively all present day, yet the root to tip path length (which is proportional to the rate of evolution) is highly dependent upon relative closeness of relationship to modern humans. Given that brain size in all these effectively present day hominins is fairly similar, and large, this suggests that, in particular, the African lineage post Neanderthal divergence, may have undergone relatively radical brain reorganization, as reflected in overall shape of the top of the skull.

Figure 16 also reinforces the conclusions with regard to the root Procrustes OLS+ tree, in terms of relationships, showing stronger resolution of all internal edges except that leading to the Neanderthals. The external edge length to LH18 is the shortest on this tree, suggesting this skull may be very close to an ancestor of the modern human readers, under this model of evolution. This accords reasonably well with its age (quite possibly 200-300kya) and location (East Africa). However, the edge leading to Dali is not much longer and its age (~200 kya) and location (China) do not accord well with an expected ancestral population to the latter samples in this tree. This tree also extends the view that the enigmatic Iwo Eleru skull most probably represents a hitherto unknown lineage and species that diverged about the time of LH18 or slightly before, and persisted in West Africa until about 12kya. That is, a lineage of pre-humans that was perhaps 150 to 400 thousand years long, apparently going extinct about 10kya as the last signs of the middle stone age (MSA) in West Africa disappeared and late stone age (LSA) technologies associated with modern humans appeared (Casey 2003).

Note, nearly all the large residuals from figure 17 are negative, which implies quite a few skulls strongly converging on others. Indeed, of the 66 residuals, three are zero due to being on tree cherries. Of the remaining 63, of these 51 are positive in sign (and relatively smaller) and only 12 are negative and relatively bigger. There is possibly a pattern here of some male skulls converging onto selected distantly related male skulls. When the residuals are sorted, those skulls showing the greatest overall sum of residuals seem to be enriched in males, while those at the bottom of the list seem enriched in females. That ordering is UMUMMUMUUFFU, which, removing the U's, yields the ordering MMMMFF. This is a perfect ordering, given the information at hand. *Homo* males are well known to develop secondary sexual characteristics, which include ridges and thickenings in areas such as above the orbits. If this is the cause, then one driver of $q$ below 1 could be the relatively rapid turn over of sexual characters on the skull due to positive sexual selection. One way to test this hypothesis in the future will be to examine specifically the shape of the inside of the neurocranium which is expected to much less influenced by sexual selection and relatively more influenced by natural selection. If this borne out, then the ordering of these residuals may offer some further clues as to the more likely sexual assignment for skulls such as LH18 and ER3883.

| | | | | | | | | | | | | | |
|---|---|---|---|---|---|---|---|---|---|---|---|---|---|
| LH18 | 0.00 | 5.08 | 0.14 | 0.03 | 2.55 | 0.17 | 2.89 | 1.28 | 0.19 | 0.21 | 2.13 | 1.81 | 16.47 |
| KABWE | 5.08 | 0.00 | 0.16 | 0.43 | 5.96 | 0.12 | 0.94 | 0.17 | 1.38 | 0.61 | 0.29 | 0.05 | 15.21 |
| ER3883 | 0.14 | 0.16 | 0.00 | 2.36 | 1.12 | 0.00 | 0.75 | 0.66 | 4.05 | 2.37 | 1.47 | 0.88 | 13.97 |
| Kh1M044H | 0.03 | 0.43 | 2.36 | 0.00 | 1.33 | 5.61 | 1.64 | 0.88 | 0.37 | 0.14 | 0.90 | 0.14 | 13.83 |
| CROMAG1 | 2.55 | 5.96 | 1.12 | 1.33 | 0.00 | 0.01 | 0.21 | 0.28 | 0.33 | 0.39 | 0.00 | 0.58 | 12.77 |
| ER3733 | 0.17 | 0.12 | 0.00 | 5.61 | 0.01 | 0.00 | 3.35 | 0.49 | 0.02 | 2.56 | 0.06 | 0.11 | 12.49 |
| CHAPELLE | 2.89 | 0.94 | 0.75 | 1.64 | 0.21 | 3.35 | 0.00 | 2.39 | 0.08 | 0.10 | 0.00 | 0.07 | 12.44 |
| IwoEleru | 1.28 | 0.17 | 0.66 | 0.88 | 0.28 | 0.49 | 2.39 | 0.00 | 1.12 | 1.01 | 0.25 | 1.09 | 9.62 |
| DALI | 0.19 | 1.38 | 4.05 | 0.37 | 0.33 | 0.02 | 0.08 | 1.12 | 0.00 | 0.10 | 0.03 | 0.10 | 7.75 |
| Kh1F043H | 0.21 | 0.61 | 2.37 | 0.14 | 0.39 | 2.56 | 0.10 | 1.01 | 0.10 | 0.00 | 0.01 | 0.00 | 7.50 |
| QUINA5 | 2.13 | 0.29 | 1.47 | 0.90 | 0.00 | 0.06 | 0.00 | 0.25 | 0.03 | 0.01 | 0.00 | 0.00 | 5.14 |
| CROMAG3 | 1.81 | 0.05 | 0.88 | 0.14 | 0.58 | 0.11 | 0.07 | 1.09 | 0.10 | 0.00 | 0.00 | 0.00 | 4.82 |

Figure 17. Squared residuals from fitting 12 taxa with $q = 1$ and $P = 2$ to the optimal tree.

Procrustes distances with Neighbor Net and all 13 taxa (figure 18) also reinforce the major features of the earlier NeighborNet of figure 7 based upon root Procrustes distances. A visual difference is that most of the external edges are much shorter, with those to Dali, LH18 and Qafzeh6 being effectively zero length. As with the 12-taxon data set, the fit of the 13 taxon data set to either tree or NeighborNet is markedly worse when $q = 1$ rather than 0.5.

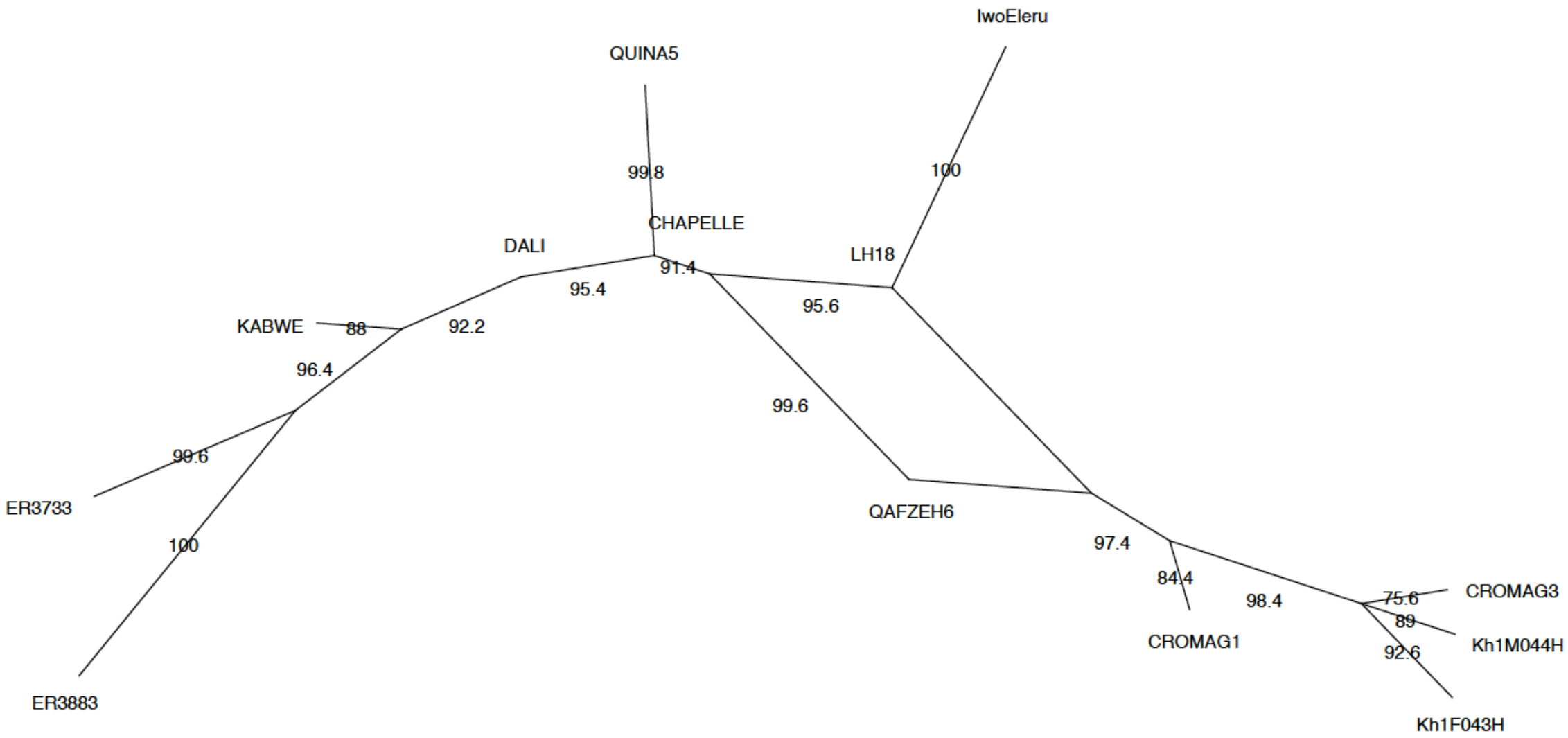

Figure 18. The planar diagram for the Procrustes distance ($q$ = 1) selected by NeighborNet (NN) with edge lengths reconstructed with the OLS+ method. The percentage of the time a split appears in replicates using residual resampling from $N(0, \hat{\sigma}^2)$ is shown, with the graph filtered to hide any split with less than 70% support. The g%SD is over twice that seen with the root Procrustes distance, so the resampling variance is over 4 times as large when adjustment is made for the scale of the distances. Four external edges, namely Dali, Chapelle, LH18 and Qafzeh6, have less than 70% support, as they often shrink to zero length on the replicates.

## 4 Discussion

The analyses presented above contain a number of strong results, which show the great promise of geometric morphometrics when coupled to more powerful distance-based phylogenetic analyses (e.g., Waddell and Azad 2009, Waddell et al. 2010, 2011a, 2011b). These results include a clearly delineated phylogenetic tree of lineages in *Homo* that largely agrees with at least a wide range of strongly voiced opinions. However, these views are not always based in solid quantitative arguments, and almost never in appropriate quantitative phylogenetic analyses (these changing views are well reviewed in Schwartz et al. 2005). For the first time, many of these views, are given clear support with appropriate quantitative phylogenetic analyses (although see also figures 9 and 10, Waddell 2013, which are largely consistent with the results here). Taxa associated with "*Homo heidelbergensis*" are shown to be paraphyletic in multiple hypothesis tests. Further, one of them (Dali) in China may show evidence of interaction with the later lineage that lead to Neanderthals. Thus, these analyses reject these "*Homo heidelbergensis*" as being directly ancestral to Neanderthals. All African skulls tentatively post 400 kya associate with the lineage leading to modern humans and after the split with Neanderthals. Iwo Eleru diverges deep and first, followed closely by LH18. Qafzeh 6 diverges about midway along the lineage from Neanderthal to the common ancestor of the KhoiSan. However, it is shown to fit a tree poorly and a NeighborNet analysis suggests strongly that it is a hybrid with about 1/3 of ancestry from close to the Neanderthal divergence and about 2/3 from well after that time and a bit before CroMag1, thus well on the way to modern humans. CroMag1 is seen to diverge consistently and markedly earlier than the other modern humans in this sample. There are no obvious hints as to why this may be.

It is important to note that Qafzeh 6, or even LH18, are sometimes described as "anatomically modern humans." The analyses in this article show how problematic such description may be as phylogenetic analyses illuminate the true situation. The correct terminology might be relatively near anatomically modern human, or, for LH 18, on the lineage towards anatomically modern humans, but not anatomically modern human if members of the population can be shown with good confidence to diverge deeper in the tree than modern humans. This echoes the concern of Schwartz and colleagues (2005, and one of the reasons for this essential book) that too many lineages are simply being called anatomically modern human, despite having features that are very rare in modern humans, or lacking derived features nearly all modern humans have. Resisting the temptation to call other lineages of *Homo* anatomically modern humans is essential in helping to clarify the origins of our own species. We are a species that most probably emerged from a country-sized area in Southern Africa only about 100kya, and with considerable prejudice and efficiency, apparently replaced/exterminated all other lineages/species of *Homo* in the world. We did this with minor (less than 10%) genetic leakage persisting into modern times. If we are to understand the profound aspects of that history and the earlier stages leading to modern humans, we need to be precise in our descriptions. Further, the view than modern humans were once endangered because they had an ancestral effective population size of ~5000-10,000 about 140kya might need to be turned on its head. It may equally be a sign of just how exclusive we were once we got going. The persistence of that relatively low number upon leaving Africa may come to be viewed as a mark of how endangered all the remaining other lineages/species of *Homo* would be after that time.

In terms of methodology, the flexi-weighted least squares (fWLS) methods (Waddell et al. 2011), now partly implemented into PAUP*, worked well. They worked especially well in combination with likelihood-based transformations of the data. A variety of analyses on phylogenetic residuals where developed, and these played an important role in discerning outliers. NeighborNet with residual resampling gave a clear picture of what seemed to be going on with Qafzeh 6, which appears quite possibly to be a hybrid. Fit criteria of the g%SD form were further developed and used, while a likelihood analogue of $R^2$, or $R^{2(L)}$, seemed to perform particularly well and integrate well with g%SD measures. Support values for phylogenetic hypotheses obtained via residual resampling, likelihood ratio tests and standard errors on edge length parameters worked well and complimented each other. With a lot of choice coming in with free or constrained parameters plus the search over tree space, model selection criteria are seen to be an important aspect that needs a lot more development. This package of methods would make a powerful addition to PAUP* and it is hoped many will be implemented in the coming year.

It is interesting to note that these large residuals from the best tree seem to bear no close relationship to convergences in shape due to size (Lieberman and Bar-Yosef 2005). For example, CroMag1 is a huge skull (probably the largest or second largest brain in this sample, Schwartz et al. 2005) while its convergence is onto ER3883, which has one of the two smallest brain sizes in this sample. This seems to apply to all the largest residuals examined in this study. In contrast, potential hybridization and convergence on certain male characteristics do seem to explain larger residuals. In the latter case it is not just being male, since not all males converge on each other. That leaves open the possibility that sexual selection in different lineages has occasionally converged on similar features, such as enlarged brows. It is probable that size effects on shape, more prevalent in archaic *Homo* (herein, Neanderthal divergence and earlier), noted by Lieberman and Bar-Yosef (2005), have been partly absorbed into the earlier internal edge lengths. However, these size effects do not seem to have disrupted the overall picture of *Homo* evolution produced by the tree and planar graphs, indeed, while Kabwe is larger than Dali, for example, it diverges markedly earlier. The results herein are consistent with Lieberman and Bar-Yosef (2005) in pointing out a substantial change certainly occurred in the lineage leading to modern humans after Neanderthal divergence, but make the novel point that it seems to be an accelerating process of shape change in that lineage that finally gave rise to humans.

The analyses here suggest no sister relationship of Iwo Eleru to LH18, except that the probable long distinct lineage to Iwo Eleru got started perhaps 200-300 kya somewhere in Africa from a form perhaps like LH18 and at a very similar time to when the lineage to LH18 diverged from the lineage leading to modern humans. LH18 was particularly close to the ancestral lineage that eventually lead on to modern humans in the Procrustes distance analyses (rather than root Procrustes analyses). Indeed the analyses suggest that the Iwo Eleru lineage may have started just a little earlier. Such a long apparently distinct lineage that terminated in West Africa perhaps 12kya, with no obvious sign of living descendants, suggests that the Iwo Eleru lineage quite probably represents a distinct species of near modern human. As such, the species name *Homo iwoelerueensis* suggests itself, being consistent with the naming of the species *Homo neanderthalensis* after the contemporary name of the location of the first clearly identified skull, as well as many other *Homo* fossil examples.

The archeological context of the Iwo Eleru burial contains important information (Shaw 1978/1979, Shaw and Daniels 1984), which has recently been reviewed (Allsworth-Jones et al. 2010). It appears to have come near the very beginning of what was a Late Stone Age (LSA) occupation, with microliths. The combination of Uranium series dates on the skeleton (12-16kya, Harvarti et al. 2012) and the nearby charcoal fragments (~11kya, Shaw and Daniels 1984) corroborate this. It was almost certainly a compact very shallow burial, between two large apparently in situ rocks, without obvious grave goods. The environment at the time of 10-16kya was probably savannah, turning into forest a few thousand years later. There is a fair amount of continuity at the site, with a shift towards ground stone tools in the forest period. The Middle Stone Age (MSA) in West Africa has now been dated from ~160 to 19kya (Casey 2003, Quickert et al. 2003, Rasse et al. 2004), and may well be associated with pre-human hominins. Further, while it was thought that modern humans might express MSA technologies with no sign of LSA technology over long periods, this view is being challenged by ever-earlier LSA industries appearing in South Africa. Whether the apparent transitions back to MSA technologies were environmental or might represent population movements is less clear. Thus there is the possibility, based on archeology alone, that Iwo Eleru represents a late survivor of the MSA in the area. West African sites such as Birimi in northern Ghana suggest a flake and Levallois industry may have existed side by side with an LSA industry, suggesting two distinct hominin populations in the area. Thus the exact level of technological achievement of the IwoEleru individual remains unclear, but the burial itself is similar to those of late Neanderthals (Rendua et al. 2014).

Thus, like the last Neanderthals, the Iwo Eleru individual comes from an area that may have been in transition from MSA to LSA. In the context of Europe, there has recently been a trend to demolish evidence for cultural transformations of Neanderthals, with LSA cultural elements after contact with modern humans, by tying all LSA materials more closely to modern humans who got into Europe much earlier than previously expected (due to nearly ubiquitous modern contamination when doing C14 dates). The same situation may hold with Iwo Eleru, with the perhaps critical difference that its population may have held a transition zone with modern humans for possibly 40,000 years or more (the time that modern humans, are expected into Central Africa based on genetics). This, combined with being phylogenetically closer to modern humans, suggests these hominins may have been both stronger competitors, but also that there might have been genetic leakage between the species. There is even some suggestion of continuity past 10kya in the region, with some ~7,000 kya West African burials showing a broad mandibular ramus like IwoEleru, but that site seems to have experienced a biological turnover in form by ~3000 kya (Ribot et al. 2001).

The existence of multiple distinct lineages of the genus *Homo* at the same time in the past remains a highly controversial topic, with some authors wanting to mix nearly all specimens prior to ~0.5 mya into *Homo erectus*, and everything since then into *Homo sapiens* (Lieberman and Bar-Yosef 2005, Lordkipanidze et al. 2013). This is emphasized by the strongly undecided ongoing debate about whether *Homo neanderthalensis* was really a distinct species or whether it should be considered a variant of modern humans. In contrast, genetics tells a story that strongly supports a multi-species view of the genus *Homo*. Modern humans seem to have originated in Southern Africa roughly 80 to 140 thousand years ago (e.g. Waddell and Penny 1996). The most distinct lineage of modern humans are the KhoiSan who show very little evidence of gene flow inwards after that separation (e.g., Waddell et al. 2011), unless relatively recently. All modern human populations, KhoiSan included, show a very rich array of mental capacities, including very symbolic behaviors. The population from which modern humans evolved was surprisingly small, and probably represents an ancestral area more like a good sized African country, not a huge continent such as Africa, or even a major region of it. There is also evidence of an early bottleneck and/or small founding population that left Southern Africa to colonize much of the rest of Africa (Waddell et al. 2011), perhaps a few tens of thousands of years before the main modern human out of Africa event which looks like roughly 60 kya.

There are multiple views of why modern human spread took so long, and seemed slow to get started, if indeed pretty much full modern mental faculties were in place at ~100kya. Aspects of this question may be enlightened by the results herein. If all modern humans, originating at least 80 kya, had mental capacities, which based on the archeological record so far, suggest at least a major quantitative (if not qualitative) leap over their predecessors and close relatives, why not immediate spread? One is physical barriers such as deserts. Indeed, these may have played a role in both the out of Southern Africa and the out of Africa delays. Another is the time required to develop new technologies/cultures to meet the challenges of changing physical environments, yet once out of Africa, indications are of relatively rapid expansion into highly varied environments with few major delays (particularly now modern humans may have colonized Europe well prior to 40kya). However, a compelling hypothesis from evolutionary theory is that competition from closely related species occupying effectively the same niche, as modern humans delayed expansion into new territory. This may be clichéd as "Hell is other people" or perhaps more accurately "Hell was other near-people", although for all these near human species, "Hell was people." Given whatever species of near modern humans existed outside of Southern Africa approximately 100 kya ago, those in the other parts of Africa were almost certainly the most similar, similarly adapted, and probably the most competitive. This includes lineages leaving evidence of probable early symbolic intent in red ochre use across Africa from ~ 300kya and possibly earlier. It was these nearest relatives that may have posed the greatest challenge to modern human expansion, with Neanderthals and relatives in second place, yet being overcome more rapidly despite their obvious advantage of adapting to very different environmental conditions for roughly 500,000 years. Indeed the rapidity with which modern humans seemed to overrun Neanderthals is surprising. They were particularly large brained and were expanding and apparently replacing other lineages of *Homo* in Eurasia starting around ~150-200kya (Bar-Yosef and Belfer-Cohen 2013). In contrast, when modern humans encountered the hobbits of Flores, they probably represented no particular challenge (except perhaps, with regards to those sinews of meat that get really stuck between your teeth).

At least one other species of near modern humans, apart from the Iwo Eleru lineage, probably existed in Africa ~100kya and these forms are particularly close to modern humans physically, but appear to be genetically and culturally distinct. These are represented by the skulls of near modern humans found in the Middle East from roughly 100kya to 60kya. Their morphology does not consistently show the modern human condition of a well-developed trigonal chin, and their archeological sites show no evidence of technology in advance of those of nearby forms that appear closely related to Neanderthals. In this analysis, the one representative of these physically near modern humans was Qafzeh 6. The analyses suggest that it arose from a lineage distinctly earlier than all the modern humans considered in this sample, yet clearly later than the lineages to LH18 and Iwo Eleru. There is also the suggestion that that particular individual may have had considerable gene flow in from yet earlier diverging lineages, with Neanderthals being an obvious choice as they were expected to be in the area at that time.

While the argument that there is evidence of considerable gene flow between distinct lineages within *Homo*, therefore we must all be the same species, is tempting, it may not be accurate. Species are not real objects, and there are multiple definitions of them. However, from an evolutionary, ecological and genetic perspective, if one lineage comprising a smaller population can invade the home territory of a numerically larger well established and locally well adapted population and largely replace it, with minimal gene exchange or introgression into the victorious population (e.g. < 20%), and with no evidence of an unprecedented ecological shift, the preponderance of the evidence must be that they were distinct species.

An interesting, and possibly profound, result of these tree analyses arises when considering the range of variation the trees present relative to those seen in non-phylogenetic analyses, such as PCA (Zelditch et al. 2004). In these non-phylogenetic analyses, “groups” or imagined species such as “near Homo sapiens”, Middle Pleistocene Homo, and Early Pleistocene hominins seem to show similar amounts of variability in skull dimensions to modern humans. This is often then interpreted to mean that in the past, that these were effectively one species with a similar range of variation to modern humans. However, the detailed analyses of the tree presented herein suggest an alternative explanation. The root Procrustes distance and even more so, the Procrustes distance tree, suggest that the rate of evolution in skull shape has increased through time, so that what may appear to be a fairly compact group around 0.7mya, for example, may indeed be several distinct lineages, each separately evolving for markedly longer than the duration of modern humans, at ~100kya.

Detailed phylogenetic analysis of these data also revealed another surprise, which was a possible reticulate association of Dali with Neanderthals. Dali has been put forward as a possible morphotype for the Denisovan individual (e.g., Reich et al. 2011). However, it has been suggested that the JinuiShan skull from China would fill this role better (Waddell 2013). However, with the results of Waddell et al. (2011), Waddell and Tan (2012) and Waddell (2013) each showing that the Denisovan had a fraction of earlier autosomal ancestry, then the situation might be interpreted slightly differently. That is, that Dali, like the Denisovan, has earlier ancestry plus ancestry closer to Neanderthals (with the fractions reversed). The fraction of more archaic ancestry in the Denisovan can jump from less than 5% to over four times as high or more if its derived alleles are from a more derived archaic than *Homo erectus*. The analyses herein suggest that the Dali skull may be just such a form. Since Dali seems to lean towards an association with Neanderthals, it suggests that in China, perhaps probable *Homo erectus* derivatives like Peking man were largely genetically replaced by a migration event even earlier than the Denisovan-lineage. That is, replaced by another member from from the dustbin-like taxon “*Homo heidelbergensis*”. Dustbin like, because these analyses clearly suggest that forms such as Dali and Kabwe, that are often grouped with *Homo heidelbergensis* (the jaw bone), may indeed be quite distinct lineages that make “*Homo heidelbergensis*” as often imagined, non-monophyletic, and indeed, probably a highly polyphyletic set of lineages.

## Acknowledgements

This work was partly supported by NIH grant 5R01LM008626 to PJW. Thanks to David Bryant, Katerina Harvati, Yunsung Kim, Hiro Kishino, Xi Tan, Jeff Schwartz, and Dave Swofford for helpful discussions and assistance with data/software/calculations. Tim Herston kindly checked some of the algebraic expressions. Special thanks to William Henry Gates for his support of the scientific community; each new version of Word seems to do a more interesting job of formatting, particularly with equations.

## Appendices

A1 The root Procrustes distances

| | | | | | | | | | | | | | |
|---|---|---|---|---|---|---|---|---|---|---|---|---|---|
| ER3733 | .0000000 | .0076921 | .0074255 | .0067293 | .0081611 | .0105933 | .0161886 | .0128358 | .0106062 | .0203070 | .0227762 | .0257233 | .0270864 |
| ER3883 | .0076921 | .0000000 | .0085763 | .0071636 | .0117390 | .0137197 | .0180694 | .0153099 | .0153254 | .0201077 | .0256958 | .0274343 | .0269452 |
| KABWE | .0074255 | .0085763 | .0000000 | .0044980 | .0073020 | .0085861 | .0123408 | .0090773 | .0091852 | .0147632 | .0192834 | .0212385 | .0185724 |
| DALI | .0067293 | .0071636 | .0044980 | .0000000 | .0036223 | .0051099 | .0105522 | .0078886 | .0067499 | .0138966 | .0161877 | .0164970 | .0172934 |
| CHAPELLE | .0081611 | .0117390 | .0073020 | .0036223 | .0000000 | .0029172 | .0076115 | .0074521 | .0065536 | .0138647 | .0151074 | .0153574 | .0170144 |
| QUINA5 | .0105933 | .0137197 | .0085861 | .0051099 | .0029172 | .0000000 | .0104069 | .0071308 | .0087691 | .0148392 | .0169812 | .0173409 | .0181283 |
| IwoEleru | .0161886 | .0180694 | .0123408 | .0105522 | .0076115 | .0104069 | .0000000 | .0064005 | .0103997 | .0132803 | .0136796 | .0140162 | .0136869 |
| LH18 | .0128358 | .0153099 | .0090773 | .0078886 | .0074521 | .0071308 | .0064005 | .0000000 | .0076853 | .0088005 | .0088381 | .0097383 | .0097895 |
| QAFZEH6 | .0106062 | .0153254 | .0091852 | .0067499 | .0065536 | .0087691 | .0103997 | .0076853 | .0000000 | .0068907 | .0067190 | .0072180 | .0079377 |
| CROMAG1 | .0203070 | .0201077 | .0147632 | .0138966 | .0138647 | .0148392 | .0132803 | .0088005 | .0068907 | .0000000 | .0061519 | .0064209 | .0052165 |
| CROMAG3 | .0227762 | .0256958 | .0192834 | .0161877 | .0151074 | .0169812 | .0136796 | .0088381 | .0067190 | .0061519 | .0000000 | .0033142 | .0037298 |
| Kh1F043H | .0257233 | .0274343 | .0212385 | .0164970 | .0153574 | .0173409 | .0140162 | .0097383 | .0072180 | .0064209 | .0033142 | .0000000 | .0038328 |
| Kh1M044H | .0270864 | .0269452 | .0185724 | .0172934 | .0170144 | .0181283 | .0136869 | .0097895 | .0079377 | .0052165 | .0037298 | .0038328 | .0000000 |

## References

Adams D.C., Rohlf F.J., Slice D.E. (2004). Geometric morphometrics: ten years of progress following the "revolution". It. J. Zool. 71: 5–16.

Adams, D.C., F. J. Rohlf, and D.E. Slice. (2013). A field comes of age: Geometric morphometrics in the 21st century. Hystrix. 24:7-14.

Agresti A. 1990. Categorical data analysis. New York: John Wiley and sons.

Akaike, H. (1974). A new look at the statistical model identification. IEEE Transactions on Automatic Control 19: 716–723.

Allsworth-Jones, P, K. Harvati, and C. Stringer. (2010). The archaeological context of the Iwo Eleru cranium from Nigeria and preliminary results of new morphometric studies. BAR International Series 2164: 29–42.

Bandelt, H.-J., and A.W.M. Dress. (1992). Split decomposition: A new and useful approach to phylogenetic analysis of distance data. Mol. Phyl. Evol. 1: 242-252.

Bar-Yosef, O., and A. Belfer-Cohen. (2013). Following Pleistocene road signs of human dispersals across Eurasia. Quaternary International 285: 30-43.

Bermúdez de Castro, J. M., M. Martinón-Torres, M. J. Sier, and L. Martín-Francés. (2014). On the Variability of the Dmanisi Mandibles. PLoS ONE 9(2): e88212. doi:10.1371/journal.pone.0088212

Brothwell, D., T. Shaw. (1971). A Late Upper Pleistocene Proto-West African Negro from Nigeria. Man, (new series), 6: 221-227.

Bryant, D., and V. Moulton. (2004). NeighborNet: an agglomerative algorithm for the construction of planar phylogenetic networks. Mol. Biol. Evol. 21:255–265.

Bryant, D., and P.J. Waddell. (1998). Rapid evaluation of least squares and minimum evolution criteria on phylogenetic trees. Mol. Biol. Evol. 15: 1346-1359.

Bulmer, M. (1991a). Use of the method of generalised least squares in reconstructing phylogenies from sequence data. Mol. Biol. Evol. 8: 868-883.

Casey, J. 2003. The Archaeology of West Africa from the Pleistocene to the Mid-Holocene. In ed. J. Mercader, *Under the Canopy: The Archaeology of Tropical Rain Forests*, 35-63. Rutgers University Press, New Jersey.

Couette, S., G. Escarguel, and S. Montuire. (2005). Constructing, bootstrapping and comparing morphometric and phylogenetic trees. Journal of Mammalogy 86(4): 773-781.

Darwin, C. (1871). The descent of Man, and selection in relation to sex. John Murray, London.

Desper, R., and O. Gascuel. (2002). Fast and accurate phylogeny reconstruction algorithms based on the minimum-evolution principle. J. Comput. Biol. 9: 687–705.

Green, R.E. et al. (2010). A draft sequence of the Neanderthal genome. Science 328: 710-715.

Felsenstein, J. (1989). PHYLIP -- Phylogeny Inference Package (Version 3.2). Cladistics 5: 164-166.

Felsenstein, J. (2004). *Inferring Phylogenies*. Sinauer Associates, Sunderland, MA.

Gascuel, O., and M. Steel. (2006). Neighbor-Joining Revealed. Molecular Biology and Evolution 23: 1997-2000.

Harvati, K, C. Stringer, R. Grün, M.Aubert, P. Allsworth-Jones and C.A. Folorunso. (2011) The Later Stone Age Calvaria from Iwo Eleru, Nigeria: Morphology and Chronology. PLoS ONE 6(9): e24024. doi:10.1371/journal.pone.0024024

Harvati, K, C. Stringer, R. Grün, M.Aubert, P. Allsworth-Jones and C.A. Folorunso. (2013) Correction: The Later Stone Age Calvaria from Iwo Eleru, Nigeria: Morphology and Chronology. PLoS ONE 8(11): 10.1371

Hendy, M.D., and D. Penny. (1993). Spectral analysis of phylogenetic data. Journal of Classification 10: 5-24.

Hillis, D. M., and J. P. Huelsenbeck. (1992). Signal, noise, and reliability in molecular phylogenetic analyses. Journal of Heredity 83: 189-195.

Huson, D.H., and D. Bryant. (2006). Application of phylogenetic networks in evolutionary studies. Mol. Biol. Evol., 23: 254-267.

Kishino, H. and M. Hasegawa. (1989). Evaluation of the maximum likelihood estimate of the evolutionary tree topologies from DNA sequence data, and the branching order in Hominoidea. J. Mol. Evol. 29: 170-179.

Klein, R. G. (2009). The Human Career: Human biological and cultural origins, 3rd edition. University of Chicago Press.

Krause, J. et al., The complete mitochondrial DNA genome of an unknown hominin from southern Siberia. Nature 464, 894 (2010).

Lieberman, D. E., and O. Bar-Yosef. (2005). Apples and Oranges: Morphological versus Behavioral Transitions in the Pleistocene. In Interpreting the Past: Essays on Human, Primate and Mammal Evolution. D.E. Lieberman, R.J. Smith, and J. Kelly, Eds. Boston: Brill Academic Publishers. Pp. 275-296.

Lordkipanidze, D., M. S. Ponce de León, A. Margvelashvili, Y. Rak, G. P. Rightmire, A. Vekua, and C. P. E. Zollikofer. (2013). A complete skull from Dmanisi, Georgia, and the evolutionary biology of early *Homo*. Science 342: 326–331.

Magori, C.C., and M. H. Day. (1983). Laetoli Hominid 18: an early *Homo sapiens* skull. Journal of Human Evolution 12: 747-753.

Millard, A. R. (2008). A critique of the chronometric evidence for hominid fossils: I. Africa and the Near East 500-50 ka. J. of Human Evolution 54: 848-874.

Mounier, A., S. Condemi, and G. Manzi. (2011). The Stem Species of Our Species: A Place for the Archaic Human Cranium from Ceprano, Italy. PLoS ONE 6:e18821.

Ota, R., P.J. Waddell, M. Hasegawa, H. Shimodaira, and H. Kishino. (2000). Appropriate likelihood ratio tests and marginal distributions for evolutionary tree models with constraints on parameters. Molecular Biology and Evolution 17: 798-803.

Penny, D., M.A. Steel, P.J. Waddell, and M.D. Hendy. (1995). Improved analyses of human mtDNA sequence support a recent African origin for Homo sapiens. Molecular Biology and Evolution 12: 863-882.

Quickert, N.A., D.I. Godfrey-Smith, J.L. Casey. (2003). Optical and thermo-luminescence dating of Middle Stone Age and Kintampo bearing sediments at Birimi, a multi-component archaeological site in Ghana. Quaternary Science Reviews 22: 1291-1297.

Rasse, M., S. Soriano, C. Tribolo, S. Stokes, E. Huysecom. (2004). La séquence Pléistocène Supérieur d'Ounjougou (Pays Dogon, Mali, Afrique de l'Ouest): Évolution géomorphologique, enregistrements sédimentaires, et changements culturels. Quaternaire, 15: 329-341.

Reich D, et al. (2010). Genetic history of an archaic hominin group from Denisova Cave in Siberia. Nature. 468: 1053-1060.

Rendua, W., et al. (2014). Evidence supporting an intentional Neandertal burial at La Chapelle-aux-Saints. PNAS 111: 81-86.

Ribot, I., R. Orban, P. de Maret. (2001). The prehistoric burials of Shum Laka rockshelter (North-West Cameroun). Musée Royal de l'Afrique Centrale, Tervuren, Belgique: Annales, Sciences Humaines: vol. 164.

Rissanen, J. (2007). "Information and Complexity in Statistical Modeling". Springer.

Schwarz, G. E. (1978). Estimating the dimension of a model. Annals of Statistics 6: 461–464.

Schwartz, J. H., I. Tattersall, R. L. Holloway, D. C. Broadfield, and M. S. Yuan. (2005). The Human Fossil Record, 4 Volume Set. Wiley-Liss.

Schwartz, J. H, I. Tattersall, and Z. Chi. (2013). Comment on "A Complete Skull from Dmanisi, Georgia, and the Evolutionary Biology of Early Homo". Science 344: 360.

Shaw, T., and S.G.H. Daniels. (1984). Excavations at Iwo Eleru, Ondo State, Nigeria. West African Journal of Archaeology, volume 14.

Shimodaira, H., Hasegawa, M., 1999. Multiple comparisons of loglikelihoods with applications to phylogenetic inference. Mol. Biol. Evol. 16, 1114–1116.

Spoor, F. (2013). Palaeoanthropology: Small-brained and big-mouthed. Nature 502: 452–453.

Stringer, C. (2012). Lone survivors: How we came to be the only humans on earth. Times Books, New York.

Stuart, A., and J.K. Ord. (1987). Kendall's advanced theory of statistics. Volume 1. 5th ed. Edward Arnold, London.

Stuart, A., and J.K. Ord. (1990). Kendall's advanced theory of statistics. Volume 2, Distribution theory. Classical inference and relationship. 5th ed. Edward Arnold, London.

Sugiura, N. (1978). Further analysis of the data by Akaike's information criterion and the finite corrections. Communications in Statistics, Theory and Methods A7, 13–26.

Swofford, D.L. (2000). Phylogenetic Analysis Using Parsimony (*and Other Methods), Version 4.0b10. Sinauer Associates, Sunderland, Massachusetts.

Swofford, D.L., G.J. Olsen, P.J. Waddell, and D.M. Hillis. (1996). Phylogenetic Inference. In: "Molecular Systematics, second edition" (ed. D. M. Hillis and C. Moritz), pp 450-572. Sinauer Assoc, Sunderland, Mass.

Trinkaus, E. (2011). Late Pleistocene adult mortality patterns and modern human establishment. Proc. Natl Acad. Sci. U S A. 108:1267-1271.

Waddell, P.J. (1995). Statistical methods of phylogenetic analysis, including Hadamard conjugations, LogDet transforms, and maximum likelihood. PhD Thesis. Massey University, New Zealand.

Waddell, P. J. (2013). Happy New Year *Homo erectus*? More evidence for interbreeding with archaics predating the modern human/Neanderthal split.. arXiv (Quantitative Biology) 1312.7749, pp 1-29.

Waddell, P.J., and A. Azad. (2009). Resampling Residuals: Robust Estimators of Error and Fit for Evolutionary Trees and Phylogenomics. arXiv 0912.5822_09, pp 1-29.

Waddell, P. J., A. Azad and I. Khan. (2011a). Resampling Residuals on Phylogenetic Trees: Extended Results. arXiv (Quantitative Biology) 1101.0020, pp 1-9.

Waddell, P. J., I. Khan, X. Tan, and S. Yoo. (2010). A Unified Framework for Trees, MDS and Planar Graphs. arXiv (Quantitative Biology) 1012.5887, pp 1-14.

Waddell, P.J., H. Kishino, and R. Ota. (2001). A phylogenetic foundation for comparative mammalian genomics. Genome Informatics Series 12: 141-154.

Waddell, P.J., H. Kishino, and R. Ota. (2002) Very Fast Algorithms for Evaluating the Stability of ML and Bayesian Phylogenetic Trees from Sequence Data. Genome Informatics 13: 82-92.

Waddell, P.J., H. Kishino and R. Ota. (2007). Phylogenetic Methodology For Detecting Protein Complexes. Mol. Biol. Evol. 24: 650-659.

Waddell, P.J., and D. Penny. (1996). Evolutionary trees of apes and humans from DNA sequences. In: "Handbook of Symbolic Evolution", pp 53-73. Ed. A.J. Lock and C.R. Peters. Clarendon Press, Oxford.

Waddell, P.J., D. Penny, M.D. Hendy, and G. Arnold. (1994). The sampling distributions and covariance matrix of phylogenetic spectra. Molecular Biology and Evolution, 11: 630-642.

Waddell, P. J., J. Ramos, and X. Tan. (2011b). *Homo denisova*, Correspondence Spectral Analysis, Finite Sites Reticulate Hierarchical Coalescent Models and the Ron Jeremy Hypothesis. arXiv (Quantitative Biology) 1112.6424, pp 1-43.

Waddell, P. J., X. Tan and I. Khan. (2010). What use are Exponential Weights for flexi-Weighted Least Squares Phylogenetic Trees? arXiv (Quantitative Biology) 1012.5882, pp 1-16.

Wolpoff, M. H. (1999). Paleoanthropology, Second Edition. McGraw Hill, New York.

Zelditch, M. L. D. L. Swiderski, H. D. Sheets, and W. L. Fink. (2004). Geometric Morphometrics for Biologists: A Primer. Elsevier Academic Press, London.

Zheng, B. and A. Agresti. (2000). Summarizing the predictive power of a generalized linear model. Statist. Med. 19: 1771–1781.